\documentclass[journal,twoside]{IEEEtran}
\usepackage[english]{babel}
\usepackage{graphicx}
\usepackage{subfigure}
\usepackage{multicol}
\usepackage{setspace}
\usepackage{amsmath}
\usepackage{epstopdf}
\usepackage{fancyhdr}
\usepackage{indentfirst}
\usepackage{enumerate}
\usepackage{caption}
\usepackage{leftidx}
\usepackage{amssymb}
\usepackage{float}
\usepackage{breqn}
\usepackage{bbm}
\usepackage{booktabs}
\usepackage{bm}

\usepackage{mathrsfs}
\usepackage{cite}
\usepackage{multirow}
\usepackage{totcount}
\usepackage{algorithm}
\usepackage{algorithmic}
\usepackage{color}
\usepackage{makecell}
\usepackage{textcomp}
\usepackage{amsthm}
\allowdisplaybreaks[4]
\begin{document}

\title{Massive MIMO-OTFS-Based Random Access for Cooperative LEO Satellite Constellations}
\author{Boxiao Shen, ~\IEEEmembership{Student Member, ~IEEE}, Yongpeng Wu, ~\IEEEmembership{Senior Member, ~IEEE}, Shiqi Gong, ~\IEEEmembership{Member, ~IEEE}, Heng Liu, ~\IEEEmembership{Member, ~IEEE}, Björn Ottersten, ~\IEEEmembership{Fellow, ~IEEE}, and Wenjun Zhang, ~\IEEEmembership{Fellow, ~IEEE}
\thanks{\emph{(Corresponding author: Yongpeng Wu)}.}
\thanks{B. Shen, Y. Wu, and W. Zhang are with the Department of Electronic Engineering, Shanghai Jiao Tong
	University, Shanghai 200240, China (e-mails:
	\{boxiao.shen, yongpeng.wu, zhangwenjun\}@sjtu.edu.cn).
}
\thanks{S. Gong and H. Liu are with the School of Cyberspace Science and Technology, and the School of Information and
	Electronics, respectively, Beijing Institute of Technology, Beijing 100081, China (e-mails: \{gsqyx, heng\_liu\_bit\_ee\}@163.com).}
\thanks{
	B. Ottersten is with the Interdisciplinary Center for Security, Reliability
	and Trust (SnT), University of Luxembourg, 1855 Luxembourg City, Luxembourg (e-mail: bjorn.ottersten@uni.lu).
}
}

\maketitle
\thispagestyle{empty}
\IEEEpeerreviewmaketitle
\begin{abstract}
This paper investigates joint device identification, channel estimation, and symbol detection for cooperative multi-satellite-enhanced random access, where orthogonal time-frequency space modulation with the large antenna array is utilized to combat the dynamics of the terrestrial-satellite links (TSLs). We introduce the generalized complex exponential basis expansion model to parameterize TSLs, thereby reducing the pilot overhead. By exploiting the block sparsity of the TSLs in the angular domain, a message passing algorithm is designed for initial channel estimation. Subsequently, we examine two cooperative modes to leverage the spatial diversity within satellite constellations: the centralized mode, where computations are performed at a high-power central server, and the distributed mode, where computations are offloaded to edge satellites with minimal signaling overhead. Specifically, in the centralized mode, device identification is achieved by aggregating backhaul information from edge satellites, and channel estimation and symbol detection are jointly enhanced through a structured approximate expectation propagation (AEP) algorithm. In the distributed mode, edge satellites share channel information and exchange soft information about data symbols, leading to a distributed version of AEP. The introduced basis expansion model for TSLs enables the efficient implementation of both centralized and distributed algorithms via fast Fourier transform. Simulation results demonstrate that proposed schemes significantly outperform conventional algorithms in terms of the activity error rate, the normalized mean squared error, and the symbol error rate. Notably, the distributed mode achieves performance comparable to the centralized mode with only two exchanges of soft information about data symbols within the constellation.

\end{abstract}


\section{Introduction} 
Internet-of-Things (IoT) is one of the critical scenarios in the next-generation communications, revolutionizing the way we live and work\cite{c32}. However, terrestrial networks cover only about 20 percent of the Earth's surface, facing substantial challenges in providing ubiquitous connectivity for diverse IoT applications, such as smart agriculture, climate monitoring and intelligent transportation systems \cite{s41}.  Low earth orbit (LEO) satellites have attracted significant research interest in the fifth generation (5G) and are expected to be pivotal for the advancement of ubiquitous connectivity in the forthcoming 6G era \cite{c29,c30,c38}. Recent efforts have focused on the deployment of satellite constellations, such as OneWeb and Starlink \cite{c31}, driven by breakthroughs in aerospace, production, and communication technologies. These developments pave the way for LEO satellite communications to offer seamless global coverage. Unlike traditional human-centric communications, IoT primarily relies on machine-type communications (MTC), characterized by bursty traffic patterns due to the sporadic activity of IoT devices \cite{c33}. In this context, the random access protocol plays a key role in supporting efficient connectivity. 

Grant-free random access (GFRA) is favored in MTC for its potential to reduce signaling overhead and power consumption while improving access capability \cite{c34,c39}. Over the past few years, many methods have been proposed for joint device identification and channel estimation (JDICE) in terrestrial GFRA systems. For instance, \cite{s33} formulated the JDICE as a sparse signal recovery problem in compressed sensing and solved it by the approximate message passing (AMP) algorithm. Alternatively, \cite{dl1} utilized a Bernoulli-Gaussian-mixture as the channel prior and adopted the vector AMP unrolled with the expectation-maximization (EM) algorithm to design a neural network that improves the performance of original vector AMP. The integration of orthogonal frequency division multiplexing (OFDM) into GFRA systems was investigated in \cite{s43}, where the AMP-based algorithm was proposed to explore the channel sparsity in the angular domain. To mitigate timing and frequency offset impacts in OFDM-based GFRA, \cite{c9} proposed to decouple and incorporate the phase shift into the measurement matrix, leading to the structured generalized AMP for JDICE. It is worth noting that these schemes in \cite{s33,dl1,s43} are tailored for block fading channels which are assumed to remain constant during one transmission, and \cite{c9} simplifies the system model by assuming that the frequency offset tends to zero.
However, the inherent high mobility of LEO satellites introduces significant Doppler shifts, leading to rapid changes in the terrestrial-satellite links (TSLs). Device movement may further contribute to considerable Doppler spread \cite{ss32}, resulting in outdated channel state information and severe inter-carrier interference. These issues will degrade the performance of existing algorithms. Consequently, current terrestrial GFRA frameworks, without adjustments or new designs, struggle to meet the demands of LEO satellite communications.

In light of the Doppler effect, extensive efforts have been devoted to GFRA adaptations for LEO satellite communications. In \cite{c36}, Bernoulli–Rician message passing combined with the EM algorithm was proposed for the LEO satellite-based narrowband GFRA, accounting for random phase rotations in the channel. However, this method still presumes a slow time-varying channel, which may be violated in dynamic satellite environments. The work in \cite{c10} focused on the OFDM-based GFRA for LEO satellite IoT, where the grid-based parametric model was proposed to discretize the delay and Doppler shift, and a message-passing type algorithm was designed for JDICE. However, the proposed scheme relies heavily on pre-compensation for both delay and Doppler shift, which limits its application in the LEO contexts. Meanwhile, \cite{c5} leveraged structured sparsity in the delay-Doppler-time domain to propose an AMP-based algorithm for joint active device, delay, and Doppler detection, while neglecting the potential non-line-of-sight paths.
Orthogonal time frequency space (OTFS) modulation has emerged as a promising solution to ensure reliable communications in high-mobility scenarios \cite{c35,c11,c4}, which converts the time-variant channels into quasi-static channels in the delay-Doppler domain \cite{c8,c15}. In \cite{c35}, OTFS is combined with tandem spreading multiple access to accommodate the differential Doppler shift characteristics in LEO satellite-enabled GFRA. Note that \cite{c36,c10,c5,c35} focuses on single-antenna systems. To exploit the spatial diversity, the integration of massive multiple-input multiple-output (MIMO) with OTFS-based GFRA has been explored in \cite{c11} and \cite{c4}. Specifically, \cite{c11} introduced a two-dimensional (2D) pattern coupled hierarchical prior within sparse Bayesian learning for JDICE, exploiting channel sparsity in the delay-Doppler-angle domain. \cite{c4} investigated OTFS aided by training sequences and proposed a two-stage JDICE scheme alongside a streamlined multi-user symbol detection method.

All the above GFRA schemes focus on one satellite node. Nevertheless, the advent of densely deployed LEO satellite constellations \cite{c17}, coupled with their inter-connectivity through highly reliable and low-latency inter-satellite links (ISLs) \cite{c3}, opens new avenues for enhanced RA schemes through cooperative processing. In \cite{c1}, the study derived closed-form expressions for spectral efficiency in ultra-dense LEO satellite networks utilizing distributed massive MIMO techniques. The theoretical analysis and simulation results, based on Rician fading with random phase modeling for TSLs, demonstrated superior performance compared to both collocated massive MIMO and conventional single-satellite connectivity. 
In addition, \cite{c2} proposed a DFT-s-OFDM framework tailored for multi-satellite cooperative GFRA with massive MIMO, which involves independent channel estimation by single satellite, with subsequent activity and data detection centralized at a high-power node.
However, \cite{c1} and \cite{c2} neglect the Doppler effect. Moreover, they are confined to the centralized cooperative mode, which requires a central server to process the collective signal, imposing huge computational burden on this server.

Overall, most previous works focus on the single satellite-based GFRA or the centralized cooperative schemes neglecting the Doppler effect. In this paper, we investigate the joint device identification, channel estimation, and symbol detection for GFRA enhanced by cooperative LEO satellite constellations with massive MIMO, where the generalized complex exponential basis expansion model (GCE-BEM) \cite{c19} is introduced as an efficient channel parameterization, and OTFS is leveraged to mitigate the doubly dispersive effect in TSLs. Beyond the centralized mode, our study examines the distributed mode, which enables the offloading of computational tasks to edge satellites. Our main contributions are summarized as follows:
\begin{itemize}
	\item The application of GCE-BEM for modeling angular domain TSLs is firstly studied. Through this parameterized channel model, we analyze the input-output relationship of the system. Then, by exploiting the 2D channel block sparsity in the angular domain, the Markov random field aided message passing algorithm with EM is designed for initial channel estimation.
	\item Centralized cooperative multi-satellite-enhanced joint device identification, channel estimation, and symbol detection are investigated. Specifically, the device identification is performed by aggregating the estimated channel energy from edge satellites. For channel estimation and symbol detection refinement, we design the approximate expectation propagation (AEP) algorithm, structured into three modules: interference cancellation module mitigates both inter-user and inter-component interference in the BEM model, providing likelihoods for channels and transmitted data symbols; channel estimation and symbol detection refinement modules utilize the soft information from the interference cancellation module to further enhance channel estimation and symbol detection.   
	\item  Distributed methods for cooperative multi-satellite-enhanced joint device identification, channel estimation, and symbol detection are investigated. Each satellite conducts the device identification by sharing the estimated channel energy. To exploit the spatial diversity of multi-satellite configurations, we propose to exchange the soft information regarding transmitted data symbols within the constellation, leading to the distributed AEP method. Simulation results indicate that only two exchanges of soft information about data symbols within the constellation are required to attain the performance comparable to that of the centralized framework.
\end{itemize}

The rest of this paper is organized as follows. Section \ref{II} introduces the system model and formulates the problem. Section \ref{III} proposes initial channel estimation algorithm.  The refinement schemes of two cooperative modes are discussed in Section \ref{IV}. Finally, Section \ref{V} evaluates the performance of proposed algorithms, followed by conclusions in Section \ref{VI}. 

\vspace{.2cm}
\emph{Notations}: The superscripts $(\cdot)^{*}$ and $(\cdot)^{\mathrm{H}}$ denote the conjugate and conjugated-transpose operations, respectively. The boldface capital letter $\mathbf X$ and lowercase letter $\mathbf x$ denote matrix and vector, respectively. $\operatorname{diag}(\mathbf{x})$ denotes the diagonal matrix with the elements of $\mathbf{x}$ on the main diagonal. $\bar{\jmath}=\sqrt{-1}$ denotes the imaginary unit. $\|\mathbf{X}\|_{\text F}$ denotes the Frobenius norm of $\mathbf{X}$, and $\|\mathbf{x}\|_{2}$ denotes the $\ell_{2}$-norm of $\mathbf{x}$. $\lceil x\rceil$ denotes the smallest integer that is not less than $x$. $\otimes$ is the Kronecker product, $\odot$ is the Hadamard product, and $\oslash$ denotes element-wise division. $(\cdot)_{M}$ denotes mod $M$. $\delta(\cdot)$ denotes the Dirac delta function. $X[a,b]$ and $x_{a,b}$ denote the $(a, b)$-th element of $\mathbf{X}$. $x[a]$ and $x_a$ denote the $a$-th element of $\mathbf{x}$. 
\begin{figure}[!htb]
	\centering
	\includegraphics[width=3.5in]{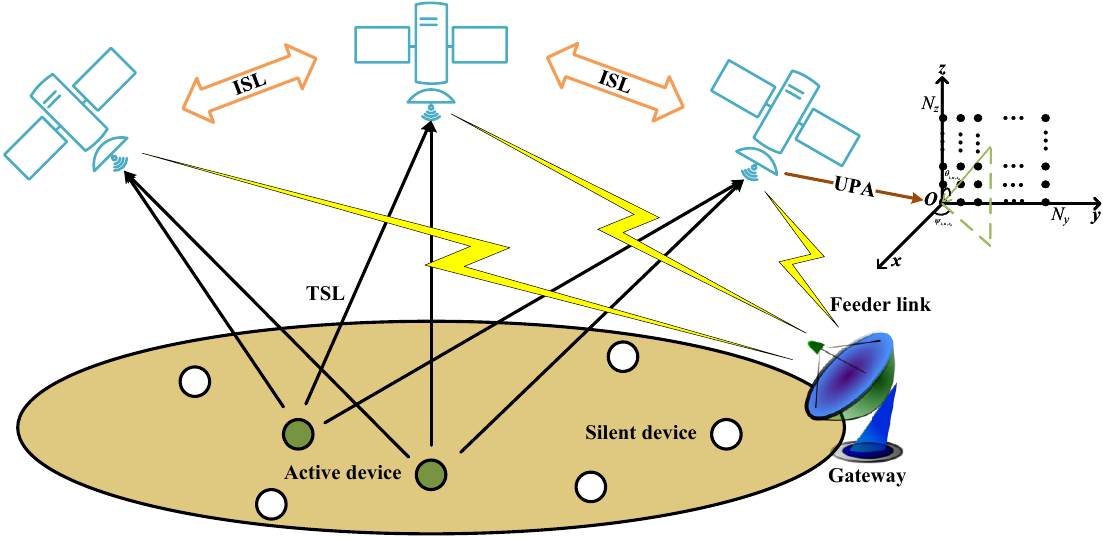}
	\caption{Diagram for cooperative multi-satellite-enhanced random access. }
	\label{system_picture}
\end{figure}

\section{System Model}
\label{II}
We investigate the LEO satellite constellations-enabled grant-free random access system as illustrated in Fig. \ref{system_picture}, where $S_a$ LEO satellites are interconnected by the high-reliable ISLs and cooperatively offer access services to $U$ single-antenna terrestrial devices. In a given time interval, active devices share the same time-frequency resources to transmit to the $S_a$ satellites. Additionally, each satellite in the constellation is equipped with a regenerative payload capable of on-board processing of baseband signals and a uniform planar array (UPA) with $N_a = N_z \times N_y$ antennas, where $N_z$ and $N_y$ are the number of antennas along the z-axis and y-axis, respectively. To mitigate the Doppler effect inherent in the TSLs, OTFS modulation is employed. Furthermore, we assume that the propagation delay can be partially pre-compensated at the devices by coarse timing advance \cite{3gpp} such that the residual delay is within one symbol duration.
\subsection{Terrestrial-satellite Link Model}
Due to the high mobility of LEO satellites, the TSLs undergo rapid variations. This variability, compounded by multipath propagation, requires a doubly dispersive channel model for accurate TSLs characterization. Consequently, the channel response for the $u$-th device to the $(n_z+n_yN_z)$-th antenna of the $s_a$-th satellite at time $n$ and tap $l^{\prime}$ is characterized as \cite{3gpp,c24,c18}
\begin{align}
\label{channel response}
	\tilde{h}_{u,s_a}^{n_z,n_y}[n, l^{\prime}]=
	\sum_{i=1}^P &h_{i,u,s_a} e^{\bar{\jmath} 2 \pi \nu_{i,u,s_a} n T_s} 
	e^{\bar{\jmath} \pi n_z \Theta_{i,u,s_a}^z} 
    e^{\bar{\jmath} \pi n_y \Theta_{i,u,s_a}^y} \nonumber \\
    &\times A_{g_{rx,tx}}\left(l^{\prime} T_s-\tau_{i,u,s_a}\right), 
\end{align}
where $u = 0,\dots,U-1$, $n_z = 0,\dots,N_z-1$, $n_y = 0,\dots,N_y-1$, $s_a = 0,\dots,S_a-1$, $P$ is the number of physical paths, and $T_s$ is the system sampling interval. The notations $h_{i,u,s_a}$, $\nu_{i,u,s_a}$, and $\tau_{i,u,s_a}$ represent the path gain, Doppler shift, and residual delay for the $i$-th path from the $u$-th device to the $s_a$-th satellite, respectively. The directional cosines $\Theta_{i,u,s_a}^z=\cos \theta_{i,u,s_a}$ and $\Theta_{i,u,s_a}^y = \sin \theta_{i,u,s_a} \sin \psi_{i,u,s_a}$, along the z-axis and y-axis, depend on the zenith angle $\theta_{i,u,s_a}$ and azimuth angle $\psi_{i,u,s_a}$, respectively. Here, we assume that the matched filter is adopted, and then the band-limited pulse shaping filter response $A_{g_{rx,tx}}\left(t\right)$ is given by
\begin{align}
	A_{g_{rx,tx}}\left(t\right) 
	= \int g_{rx}^*\left(t^{\prime} - t\right) g_{tx}(t^{\prime}) dt^{\prime},
\end{align} 
where $g_{tx}\left(t\right)$ and $g_{rx}\left(t\right)$ are the transmit and receive waveform, respectively. Contrary to prior OTFS studies \cite{c8,c11,c4}, which often require Doppler shifts or delays to be integer multiples of the frequency and delay resolutions, our framework does not rely on this assumption. The channel model in (\ref{channel response}) accommodates fractional Doppler shifts and delays, making it more suited for realistic satellite communication scenarios. Moreover, it's reasonable to assume that the propagation paths from a single device to a satellite have identical angles due to the satellite's significantly higher altitude compared to that of scatterers near the devices \cite{ss32}, i.e., $\theta_{i,u,s_a}=\theta_{u,s_a}$ and $\psi_{i,u,s_a} = \psi_{u,s_a}$. Then, the channel model in (\ref{channel response}) can be simplified as
\begin{align}
\label{channel response_simplified}
&\tilde{h}_{u,s_a}^{n_z,n_y}[n, l^{\prime}]=
e^{\bar{\jmath} \pi n_z \Theta_{u,s_a}^z} 
e^{\bar{\jmath} \pi n_y \Theta_{u,s_a}^y} 
\sum_{i=1}^P h_{i,u,s_a} e^{\bar{\jmath} 2 \pi \nu_{i,u,s_a} nT_s} \nonumber \\
&\qquad \qquad\qquad\times A_{g_{rx,tx}}\left(l^{\prime} T_s-\tau_{i,u,s_a}\right). 
\end{align}
It is observed that the term $e^{j 2 \pi \nu_{i,u,s_a} nT_s}$ exhibits significant temporal variation due to the Doppler effect, in contrast to the other terms which remain relatively constant over time. This fact motivates us to adopt the GCE-BEM \cite{c19} to parameterize this time varying channel as 
\begin{align}
\label{BEMmodel}
	\tilde{h}_{u,s_a}^{n_z,n_y}[n, l^{\prime}]=
	e^{\bar{\jmath} \pi n_z \Theta_{u,s_a}^z} 
	e^{\bar{\jmath} \pi n_y \Theta_{u,s_a}^y} 
	\sum_{q=0}^Q \tilde{c}_{q, l^{\prime}, u, s_a} e^{\bar{\jmath} \omega_q n},
\end{align}
where $Q$ is the order of the BEM basis functions, 
$\tilde{c}_{q, l^{\prime}, u, s_a}$ is the $q$-th BEM coefficient of $l^{\prime}$-th channel tap from the $u$-th device to the $s_a$-th satellite, and $\omega_q$ is the $q$-th BEM modeling frequency quantizing the Doppler shift.
Note that the modeling error term is omitted in (\ref{BEMmodel}) primarily since we focus on BEMs that provide an excellent fit. As highlighted in \cite{c37}, including the modeling error term does not significantly improve the reliability of channel estimates if the BEM itself cannot accurately characterize the true channel. Furthermore, \cite{c20} shows that, under certain mild conditions, the modeling error can be approximated as a Gaussian random variable with zero mean, allowing it to be absorbed into the noise term. To exploit the sparsity of channel resulted from adoption of the massive antenna array, the two dimensional discrete Fourier transform (DFT) is applied for transforming the channel from space domain into angular domain, i.e.,
\begin{align}
\label{ha}
	h_{u,s_a}^{a_z,a_y}[n, l^{\prime}]&=
	\sum_{n_z=0}^{N_z-1}\sum_{n_y=0}^{N_y-1}
	\tilde{h}_{u,s_a}^{n_z,n_y}[n, l^{\prime}]
	e^{-\bar{\jmath} 2 \pi (\frac{a_z n_z}{N_z}+\frac{a_y n_y}{N_y})} \nonumber \\
	&=\sum_{q=0}^Q c_{q, l^{\prime}, u, s_a}^{a_z,a_y} e^{\bar{\jmath} \omega_q n},
\end{align}
where $a_z = 0,\dots,N_z-1$ and $a_y = 0,\dots,N_y-1$. The parameter $c_{q, l^{\prime}, u, s_a}^{a_z,a_y} = \tilde{c}_{q, l^{\prime}, u, s_a}\Pi_{N_z}(a_z-N_z\Theta_{u,s_a}^z/2) \Pi_{N_y}(a_y-N_y\Theta_{u,s_a}^y/2)$ is the BEM coefficient in the angular domain, where $\Pi_N(x)\triangleq\frac{1}{N} \sum_{i=0}^{N-1} e^{-\bar{\jmath} 2 \pi \frac{x}{N} i}$. It is observed that $c_{q, l^{\prime}, u, s_a}^{a_z,a_y}$ has the dominant element only if  $a_z \approx N_z\Theta_{u,s_a}^z/2$ and $a_y \approx N_y\Theta_{u,s_a}^y/2$, which means that the coefficient has 2D block sparsity in the angular domain. For notation simplicity, we denote the index of $n_a$-th antenna as $n_a = a_z + a_yN_z$, $n_a=0,\dots,N_zN_y-1$, and then (\ref{ha}) can be re-expressed as 
\begin{align}
	\label{has}
	h_{u,s_a}^{n_a}[n, l^{\prime}]
	=\sum_{q=0}^Q c_{q, l^{\prime}, u, s_a}^{n_a} e^{\bar{\jmath} \omega_q n}.
\end{align}

\subsection{Transmission Signal Model}

In this subsection, the transmission process is explained. Each terrestrial device employs OFDM-based OTFS modulation. Firstly, the $u$-th device rearranges a sequence of $MN$ information symbols into one OTFS frame $\mathbf{X}^{\mathrm{DD}}_u \in \mathbb{C}^{M \times N}$, where $M$ and $N$ are the number of subcarriers and OFDM symbols within one frame, respectively. Then, by the inverse symplectic finite Fourier transform (ISFFT)\cite{c15}, the 2D data frame $\mathbf{X}^{\mathrm{DD}}_u$ in the delay-Doppler domain is mapped to $\mathbf{X}^{\mathrm{TF}}_u$ in the time-frequnecy domain, i.e.,
\begin{align}
\label{modulationstart}
	\mathbf{X}^{\text{TF}}_u =\mathbf{F}_{M} \mathbf{X}^{\mathrm{DD}}_u \mathbf{F}_{N}^{\mathrm{H}},
\end{align} 
where $\mathbf{F}_{\mathrm{M}} \in \mathbb{C}^{M \times M}$ and $\mathbf{F}_{\mathrm{N}} \in \mathbb{C}^{N \times N}$ are normalized DFT matrices. The OFDM modulator processes the 2D data frame $\mathbf{X}^{\mathrm{TF}}_u$ to get the transmitted data frame $\mathbf{B}_u^{\mathrm{TD}}$, i.e.,
\begin{align}
\label{B}
	\mathbf{B}_u^{\mathrm{TD}}= {\mathbf{F}_{M}^{\mathrm{H}} \mathbf{X}^{\mathrm{TF}}_u} = \mathbf{X}^{\mathrm{DD}}_u \mathbf{F}_N^H.
\end{align}
Then, the data frame $\mathbf{B}_u^{\mathrm{TD}}$ is transformed to the 1D transmit signal $\mathbf{b}_u^{\mathrm{TD}} = \operatorname{vec}\left(\mathbf{B}_u^{\mathrm{TD}}\right)$,
where $\operatorname{vec}(\cdot)$ denotes the vectorization operator stacking the columns of one matrix into a long column
vector. Note that in the traditional OTFS literature \cite{c11,c4}, the guard interval will be added for each OFDM symbols to avoid inter-symbol interference. Here, we only insert one cyclic prefix (CP) with length of $M_{\text{cp}}$ before the head of the OTFS frame for increasing spectral efficiency. Typically, $M_{\text{cp}}$ should be more than the largest normalized residual delay $L=\lceil \tau_{\text{max}}/T_s \rceil$, where $\tau_{\text{max}}$ represents the largest residual delay among all the resolvable physical paths in TSLs.
After CP removal, the received signal is given by
\begin{align}
\label{r}
	r_{u,s_a}^{n_a}[n]=\sum_{l^{\prime}=0}^L h_{u,s_a}^{n_a}\left[n+M_{\text{cp}}, l^{\prime}\right] b_u^{\mathrm{TD}}\left[\left(n-l^{\prime}\right)_{M N}\right],
\end{align} 
where $n=0,\dots,MN-1$ and noise is neglected here for simplicity. Then, this sequence of data $\mathbf{r}_{u,s_a}^{n_a}=[	r_{u,s_a}^{n_a}[0],\dots,r_{u,s_a}^{n_a}[MN-1]]$ is rearranged to the 2D data frame $\mathbf{R}_{u,s_a}^{n_a} = \operatorname{vec}^{-1}(\mathbf{r}_{u,s_a}^{n_a})$, where $\operatorname{vec}^{-1}(\cdot)$ is the inverse operation of $\operatorname{vec}(\cdot)$. Applying the $M$-point DFT on each OFDM symbol, we can obtain the received signal in the time-frequency domain as 
\begin{align}
	\mathbf{Y}_{n_a,u,s_a}^{\mathrm{TF}}=\mathbf{F}_{\mathrm{M}} \mathbf{R}_{u,s_a}^{n_a}.
\end{align}
Finally, the data frame $\mathbf{Y}_{n_a,u,s_a}^{\mathrm{TF}}$ is transformed into $\mathbf{Y}^{\text{DD}}_{n_a,u,s_a}$ in the delay-Doppler domain, i.e.,
\begin{align}
\label{YDD}
	\mathbf{Y}^{\text{DD}}_{n_a,u,s_a} = \mathbf{F}_{M}^{\mathrm{H}} \mathbf{Y}_{n_a,u,s_a}^{\mathrm{TF}} \mathbf{F}_{N}
	= \mathbf{R}_{u,s_a}^{n_a} \mathbf{F}_{N}.
\end{align}
We denote the $(l,k)$-th element of $\mathbf{Y}^{\text{DD}}_{n_a,u,s_a}$ and $\mathbf{X}^{\text{DD}}_u$ as $Y^{\text{DD}}_{n_a,u,s_a}[l,k]$ and $X^{\text{DD}}_u[l,k]$ respectively, where $l=0,\dots,M-1$ and $k=0,\dots,N-1$. According to (\ref{has})-(\ref{YDD}), the input-output relationship  (without noise term) of this system is given in the following proposition.

\textit{Proposition 1}: Given the $q$-th BEM modeling frequency $\omega_q = \frac{2 \pi q^{\prime}}{M N R}$ with $q^{\prime} = \left(q-\left\lceil\frac{Q}{2}\right\rceil\right)$ and the resolution factor $R$, the received signal in the delay-Doppler-angle domain from the $u$-th device to the $n_a$-th antenna of the $s_a$-th satellite is represented as

\begin{align}
\label{YDDs}
	Y^{\text{DD}}_{n_a,u,s_a}[l,k]=
	\sum_{q=0}^Q \sum_{l^{\prime}=0}^L  \sum_{k^{\prime}=0}^{N-1} 
	&e^{\bar{\jmath} \omega_q l}
	\alpha_{l, l^{\prime}}\left[k, k^{\prime}, q^{\prime}\right]
	c_{q, l^{\prime}, u, s_a}^{n_a} \nonumber \\
	&\times X^{\text{DD}}_u\left[\left(l-l^{\prime}\right)_M, k^{\prime}\right],
\end{align} 
where we have
\begin{align}
	\alpha_{l, l^{\prime}}\left[k, k^{\prime}, q^{\prime}\right]= \begin{cases}
	\frac{1}{N} \frac{1-e^{\bar{\jmath} 2 \pi\left(k^{\prime}-\left(k-\frac{q^{\prime}}{R}\right)\right)}}{1-e^{\bar{\jmath} \frac{2 \pi}{N}\left(k^{\prime}-\left(k-\frac{q^{\prime}}{R}\right)\right)}}, & l^{\prime} \leqslant l \\ 
	\frac{1}{N} e^{-\bar{\jmath} \frac{2 \pi}{N} k^{\prime}} \frac{1-e^{\bar{\jmath} 2 \pi\left(k^{\prime}-\left(k-\frac{q^{\prime}}{R}\right)\right)}}{1-e^{\bar{\jmath} \frac{2 \pi}{N}\left(k^{\prime}-\left(k-\frac{q^{\prime}}{R}\right)\right)}}, & l^{\prime}>l \end{cases}. \nonumber
\end{align}

\emph{Proof}: Please see Appendix \ref{App1}. $\hfill\blacksquare$

It is observed that the received signal in the delay-Doppler-angle domain is given by the sum of $Q$ components. Each component is the result of phase compensated periodic convolution between the OTFS data frame and the BEM coefficient. Here, the order $Q$ and the resolution factor $R$ govern the modeling accuracy of the BEM. At $R=1$, the GCE-BEM is simplified to the CE-BEM which has higher modeling error. Increasing $R$ to $2$ enhances the Doppler shift quantization precision, thereby significantly reducing the modeling error. As indicated in \cite{c19}, $R=2$ is large enough for modeling time-varying channel, with only marginal gain from further increases. Additionally, the order $Q$ usually depends on $R$, given as $Q \geq 2\left\lceil R N \bar{\nu}_{\max}\right\rceil$, where $\bar{\nu}_{\max}=\nu_{\max}/\Delta f$; $\nu_{\max}$ is the largest Doppler shift across all the resolvable physical paths in TSLs and $\Delta f$ is the subcarrier spacing.  Under the BEM modeling, the unknown channel parameters from the $u$-th device to the $s_a$-th satellite have been reduced significantly from $MN(L+1)$ to $(Q+1)(L+1)$ at each antenna. However, a higher resolution leads to an increased BEM order, which may in turn raise the interference among the $Q$ components in (\ref{YDDs}). This interference makes it difficult to accurately estimate channel and detect transmitted symbols. Therefore, we should make $Q$ as small as possible while ensuring the BEM modeling accuracy. 
\vspace{-0.4cm}
\subsection{Problem Formulation}
According to Proposition 1 and superposition principle, the received signal at the $n_a$-th antenna of the $s_a$-th satellite is expressed as
\begin{align}
	\label{YDDsa}
	Y^{\text{DD}}_{n_a,s_a}[l,k]=
	&\sum_{u=0}^{U-1}\sum_{q=0}^Q \sum_{l^{\prime}=0}^L  \sum_{k^{\prime}=0}^{N-1} 
	\lambda_u
	c_{q, l^{\prime}, u, s_a}^{n_a} 
	e^{\bar{\jmath} \omega_q l}
	\alpha_{l, l^{\prime}}\left[k, k^{\prime}, q^{\prime}\right] \nonumber \\
	&\times X^{\text{DD}}_u\left[\left(l-l^{\prime}\right)_M, k^{\prime}\right] 
	+ E_{s_a}^{n_a}[l,k],
\end{align}
where $\lambda_u\in \{0,1\}$ is the activity indicator of the $u$-th device with active probability $p_{\lambda}$ and we denote the set of active devices as $\mathcal{U}_a=\{u\mid \lambda_u=1, u=0,\dots,U-1\}$; $E_{s_a}^{n_a}[l,k]\sim \mathcal{CN}(0,\sigma^2_{s_a})$ is the noise. To facilitate algorithm design, we rewrite (\ref{YDDsa}) as the matrix form, i.e.,
\begin{align}
\label{Ymat}
	\mathbf{y}_{s_a}^{n_a}&=\sum_{u=0}^{U-1}\sum_{q=0}^Q
	\left(\mathbf{F}_N \otimes \mathbf{I}_M\right)
	\operatorname{diag}\left(\mathbf{b}_q\right) 
	\mathbf{F}_{MN}^{\mathrm{H}} \times \nonumber \\  
	&\operatorname{diag}\left(\mathbf{F}_{1:L+1} \mathbf{c}_{q,u,s_a}^{n_a}\right) \mathbf{F}_{MN}
	\left(\mathbf{F}_N^{\mathrm{H}} \otimes \mathbf{I}_M\right) 
	\mathbf{x}_u + \mathbf{e}_{s_a}^{n_a},
\end{align}
where $\mathbf{y}_{s_a}^{n_a} = \operatorname{vec}(\mathbf{Y}^{\text{DD}}_{n_a,s_a})$,   
$\mathbf{x}_u=\operatorname{vec}(\mathbf{X}^{\text{DD}}_{u})$, $\mathbf{e}_{s_a}^{n_a} = \operatorname{vec}(\mathbf{E}_{s_a}^{n_a})$, $\mathbf{I}_M \in \mathbb{R}^{M \times M}$ is the identity matrix,
$\mathbf{b}_q=[1,e^{\bar{\jmath}\omega_q}, \dots, e^{\bar{\jmath}\omega_q(MN-1)}]$ is the $q$-th BEM basis function, $\mathbf{F}_{1:L+1}$ represents the first $L+1$ columns of $\mathbf{F}_{MN}$, and $\mathbf{c}_{q,u,s_a}^{n_a}=\sqrt{MN}\lambda_u[c_{q, 0, u, s_a}^{n_a},\dots,c_{q, L, u, s_a}^{n_a}]$ collecting the $q$-th BEM coefficients for $L$ channel taps and combining the device activity indicator.

Given all the received signals of LEO satellite constellations, we aim at jointly estimating $\left(\mathbf{c}_{q,u,s_a}^{n_a}, \mathbf{x}_u \right)$ based on the maximum a posteriori probability (MAP) principle, i.e.,
\begin{align}
	\label{MAP}
	(\mathbf{\hat c}_{q,u,s_a}^{n_a}, \mathbf{\hat x}_u)
	=\underset{(\mathbf{c}_{q,u,s_a}^{n_a}, \mathbf{x}_u)}{\text{arg max}} p\left(\mathbf{c}_{q,u,s_a}^{n_a}, \mathbf{x}_u \mid \mathbf y \right),
\end{align}
where $\mathbf y = [\mathbf{y}_{0}^{0},\dots,\mathbf{y}_{s_a}^{n_a},\dots,\mathbf{y}_{S_a}^{N_zN_y-1}]$. The problem (\ref{MAP}) is generally non-convex and difficult to solve. Message passing algorithms offer potential solutions, but the coupling of variables into a bilinear function complicates the implementation of exact belief propagation (BP) based on the sum-product rule. Furthermore, the accuracy of bilinear inference problems is highly dependent on the initial values of the solution variables \cite{c26}, necessitating a reasonable initialization method for precise channel and data estimation. To overcome these issues, we propose a two-phase algorithm. The first phase based on the GAMP \cite{sss} algorithm to generate initial estimates of the channel and transmitted symbols. The second phase resorts to the expectation propagation (EP) \cite{c28} algorithm to enhance device identification, channel estimation, and symbol detection.
\section{Initial Channel Estimation for Single Satellite}
\label{III} 
In this section, we firstly illustrate the OTFS data frame structure. Then, based on the Bayesian method, a message-passing type algorithm combined with EM is designed for the initial channel estimation. For sake of low computations and pilot overhead, similar to \cite{c20}, we adopt a lower resolution $R=1$ with the smaller order of BEM $Q_{\mathrm I} \geq 2\left\lceil N \bar{\nu}_{\max}\right\rceil$ in this initial phase, and higher resolution is utilized in the refine phase. 
\vspace{-0.5cm}
\subsection{Frame Structure}
\begin{figure}[!htb]
	\centering
	\includegraphics[width=2.5in]{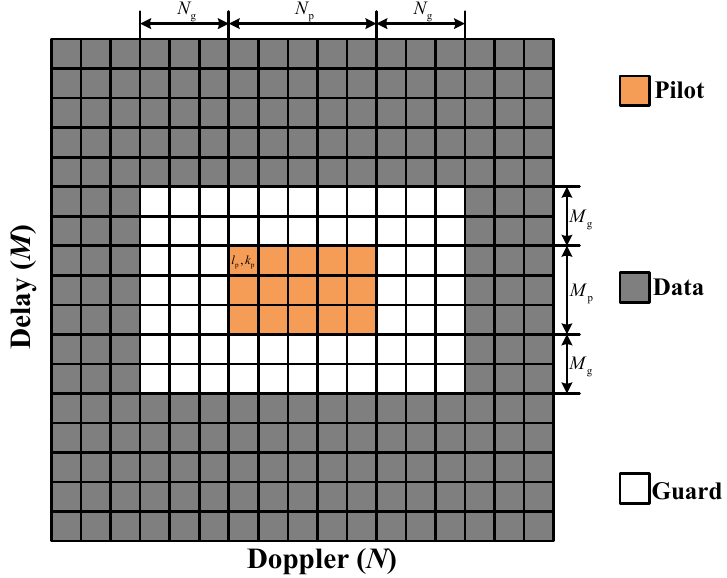}
	\caption{OTFS frame structure. }
	\vspace{-0.2cm}
	\label{frame_structure}
\end{figure}
When $R=1$, (\ref{YDDsa}) can be simplified as 
\begin{align}
	Y^{\text{DD}}_{n_a,s_a}[l,k]&=\sum_{u=0}^{U-1}
	\sum_{q=0}^{Q_{\mathrm I}} \sum_{l^{\prime}=0}^L  
	\lambda_u e^{\bar{\jmath} \omega_q l}
	\alpha_{l, l^{\prime}}\left[k, q^{\prime}\right]
	c_{q, l^{\prime}, u, s_a}^{n_a} \nonumber \\
	&\times X^{\text{DD}}_u\left[\left(l-l^{\prime}\right)_M, (k-q^{\prime})_N\right] + E_{s_a}^{n_a}[l,k],
\end{align}
where
\begin{align}
	\alpha_{l, l^{\prime}}\left[k, q^{\prime}\right]= \begin{cases}
		1, & l^{\prime} \leqslant l \\ 
		 e^{-\bar{\jmath} \frac{2 \pi}{N} (k-q^{\prime})}, & l^{\prime}>l \end{cases}.
\end{align}
By considering this 2D periodic convolution relationship, the OTFS frame structure in the delay-Doppler domain is depicted in Fig. \ref{frame_structure}. The position of pilots is given by $l=l_{\mathrm{p}}, \cdots, l_{\mathrm{p}}+M_{\mathrm{p}}-1$ and $k=k_{\mathrm{p}}, \cdots, k_{\mathrm{p}}+N_{\mathrm{p}}-1$, where $(l_{\mathrm{p}}, k_{\mathrm{p}})$ is the initial position of pilots. $M_{\mathrm{p}}$ and $N_{\mathrm{p}}$ are the length of pilots along the delay and Doppler dimension, respectively. Without loss of generality, we assume that $l_{\mathrm p} \geq L$. To avoid interference between pilots and data, guard intervals along delay and Doppler dimensions with length of $M_{\mathrm g} \geq L$ and $N_{\mathrm{g}} \geq \left\lceil\frac{Q_{\mathrm I}}{2}\right\rceil$ are required. Additionally, the pilots and guard symbols are the i.i.d. complex Gaussian random sequences. Note that guard symbols can be considered as one kind of pilots. Therefore, the pilot overhead of one OTFS frame is obtained as $\rho = \frac{(N_{\mathrm{p}}+2N_{\mathrm{g}})(M_{\mathrm{p}} + 2M_{\mathrm{g}})}{MN}$.

Given the OTFS frame structure in Fig. \ref{frame_structure}, the received symbols with the same positions as pilots are utilized for channel estimation for each satellite and can be written as
\begin{align}
	\label{Yp1}
	\mathbf{Y}_{s_a} = \left((\mathbf{1}_U^{\mathrm T} \otimes \mathbf{P}^{\omega}) \odot \mathbf{X}^{\mathrm p}\right)\mathbf{C}_{s_a} + \mathbf{E}_{s_a}^{\mathrm p},
\end{align} 
where $\mathbf{Y}_{s_a} \in  \mathbb{C}^{M_{\mathrm p}N_{\mathrm p} \times N_zN_y}$ with the $(l-l_{\mathrm p}+k-k_{\mathrm p}, n_a)$-th element $Y^{\text{DD}}_{n_a,s_a}[l,k]$, $\mathbf{1}_U\in\mathbb{C}^{U}$ is the all-one vector, and $\mathbf{P}^{\omega} \in \mathbb{C}^{M_{\mathrm{p}}N_{\mathrm{p}} \times (Q_{\mathrm I}+1)(L+1)}$ with the $(l-l_{\mathrm p}+\tilde{l}M_{\mathrm{p}}, q+\bar{l})$-th element $e^{\bar{\jmath} \omega_q l}$, $\tilde{l}=0,\dots,N_{\mathrm{p}}-1$, $\bar{l}=0,\dots,L$. $\mathbf{X}^{\mathrm p} = [\mathbf{X}^{\mathrm p}_0,\dots,\mathbf{X}^{\mathrm p}_{U-1}]$ contains the pilots of $U$ devices, where $\mathbf{X}^{\mathrm p}_u \in \mathbb{C}^{M_{\mathrm{p}}N_{\mathrm{p}} \times (Q_{\mathrm I}+1)(L+1)}$ with the $(l-l_{\mathrm{p}} + (k-k_{\mathrm{p}})M_{\mathrm{p}}, l^{\prime}+q(L+1))$-th element $X^{\text{DD}}_u\left[\left(l-l^{\prime}\right)_M, (k-q^{\prime})_N\right]$, $\mathbf{C}_{s_a} \in \mathbb{C}^{U(Q_{\mathrm I}+1)(L+1)\times N_zN_y}$ with the $(u(Q_{\mathrm I}+1)(L+1)+q(L+1)+l^{\prime}, n_a)$-th element $\lambda_u c_{q, l^{\prime}, u, s_a}^{n_a}$, and $\mathbf{E}_{s_a}^{\mathrm p}$ is the Gaussian noise with the $(l-l_{\mathrm p}+k-k_{\mathrm p}, n_a)$-th element $E_{s_a}^{n_a}[l,k]$. With the definition $\mathbf X^{\omega} \triangleq ((\mathbf{1}_U^{\mathrm T} \otimes \mathbf{P}^{\omega}) \odot \mathbf{X}^{\mathrm p})$, we rewrite (\ref{Yp1}) as 
\begin{align}
	\label{Yp2}
	\mathbf{Y}_{s_a} = \mathbf X^{\omega} \mathbf{C}_{s_a} + \mathbf{E}_{s_a} 
	= \mathbf Z_{s_a} + \mathbf{E}^{\mathrm p}_{s_a},
\end{align}
where $\mathbf Z_{s_a}\triangleq \mathbf X^{\omega} \mathbf{C}_{s_a}$. Then, the channel estimation turns into a sparse signal reconstruction problem. Next, the compressed sensing-based algorithm is designed for channel recovery, and simultaneously, exploiting the 2D block sparsity of the channel in the angular domain. 
\subsection{Initial Channel Estimation Algorithm}
\subsubsection{Prior distribution}
To estimate the channel in (\ref{Yp2}), we resort to Bayesian method which needs prior distribution of the estimated variables. Firstly, we model the channel as $K$-components Bernoulli-Gaussian-mixture (BGM) distribution, which is a general and accurate distribution for the practical scenarios with a large-scale antenna array \cite{dl1}. Additionally, we assign different BGM distributions for the channel of different devices, i.e.,
\begin{align}
p\left(c^{u,s_a}_{i, j} \mid s^{u,s_a}_{i, j}\right)&=\delta\left(s^{u,s_a}_{i, j}+1\right) \delta\left(c^{u,s_a}_{i, j}\right)+\delta\left(s^{u,s_a}_{i, j}-1\right) \nonumber \\
&\times \sum_{k=1}^K \omega_k^{u,s_a} \mathcal{C N}\left(c^{u,s_a}_{i, j} \mid \mu_k^{u,s_a}, \phi_k^{u,s_a}\right),	
\end{align}
where $c^{u,s_a}_{i, j}$ is the $(u(Q_{\mathrm I}+1)(L+1)+i, j)$-th element of $\mathbf{C}_{s_a}$, $i=0,\dots,(Q_{\mathrm I}+1)(L+1)-1$, $j=0,\dots,N_zN_y-1$, corresponding to the channel of the $u$-th device, $s^{u,s_a}_{i, j}\in\{+1,-1\}$ is the corresponding support, and $\{\omega_k^{u,s_a}, \mu_k^{u,s_a}, \phi_k^{u,s_a}\}$ are the parameters of the BGM distribution. Furthermore, we adopt the Markov random field (MRF) prior to exploit 2D block sparsity of the channel, and then its support can be modeled by the classic Ising model as
\begin{align}
	\label{support}
	p(\mathbf s^{u,s_a}_{i}) 
	=& \left(\prod_{j=0}^{N_zN_y-1} \prod_{j^{\prime} \in \mathcal{D}_j} \Gamma\left(s^{u,s_a}_{i,j}, s^{u,s_a}_{i,j^{\prime}}\right)\right)^{\frac{1}{2}} \nonumber \\
	&\times \prod_{j=0}^{N_zN_y-1} \Psi\left(s^{u,s_a}_{i, j}\right),
\end{align} 
where $\mathbf s^{u,s_a}_{i}=[s^{u,s_a}_{i,0},\dots,s^{u,s_a}_{i,N_zN_y-1}]$ involves the supports of $(u(Q_{\mathrm I}+1)(L+1)+i)$-th row in $\mathbf{C}_{s_a}$, $\mathcal{D}_j \subset \{0,\dots,N_zN_y-1\}\backslash j$ contains the indexes of the neighbors of index $j$, $\Gamma(s^{u,s_a}_{i,j}, s^{u,s_a}_{i,j^{\prime}}) = \exp(\beta s^{u,s_a}_{i, j} s^{u,s_a}_{i,j^{\prime}})$, and $\Psi(s^{u,s_a}_{i, j})=\exp(- \gamma s^{u,s_a}_{i, j})$. $\beta$ and $\gamma$ are the parameters of
MRF prior, where a larger $\beta$
implies a larger size of each block of nonzeros, and a larger $\gamma$ encourages a sparser $\mathbf{C}_{s_a}$. 
Based on (\ref{Yp2})-(\ref{support}), the posterior distribution related to channel is given by
\begin{align}
	\label{pc1}
	p(\mathbf Z_{s_a}, &\mathbf C_{s_a}, \mathbf S_{s_a} \mid \mathbf Y_{s_a})\propto \prod_{u} \prod_{i} p(\mathbf s_{i}^{u,s_a}) \prod_{j} p\left(c^{u,s_a}_{i, j} \mid s^{u,s_a}_{i, j}\right) \nonumber \\
	&\prod_{m} \prod_{j} p\left(y_{m, j}^{s_a} \mid z_{m,j}^{s_a}\right)\delta\left(z_{m,j}-\mathbf c_{{s_a}_j}^{\mathrm T} \mathbf x^{\omega}_m \right), 	
\end{align}
where $y_{m, j}^{s_a}$ and $z_{m,j}^{s_a}$ are the $(m,j)$-th element of $\mathbf Y_{s_a}$ and $\mathbf Z_{s_a}$, respectively. $\mathbf c_{{s_a}_j}$ is the $j$-th column of $\mathbf C_{s_a}$ and $\mathbf x^{\omega}_m$ is a column vector containing the elements of $m$-th row of $\mathbf X^{\omega}$. Unfortunately, the MMSE or MAP estimation of $\mathbf C_{s_a}$ with respect to (\ref{pc1}) is still hard to carry out straightforwardly, since it involves marginalizing a joint distribution with high dimensions. 
In order to obtain a tractable alternative, the GAMP algorithm is firstly leveraged to eliminate inter-device interference in (\ref{Yp2}) and obtain the scalar likelihood for $c^{u,s_a}_{i, j}$. Then the message passing algorithm is combined with GAMP to get the approximated MMSE estimation of channel. 

\subsubsection{MMSE estimation}
We denote $f^{c^{u,s_a}}_{i,j}$ as the factor corresponding to $p(c^{u,s_a}_{i, j} \mid s^{u,s_a}_{i, j})$; $\Delta^m_n$ and $\Delta_m^n$ represent the message passed from node $m$ to $n$ and the message in the corresponding  opposite direction, respectively. Now we focus on the message passing related to the MRF prior. Note that the message $\Delta_{f^{c^{u,s_a}}_{i,j}}^{c^{u,s_a}_{i,j}}$ is approximated as the Gaussian distribution by GAMP, given as
\begin{align}
	\label{likec}
	\Delta_{f^{c^{u,s_a}}_{i,j}}^{c^{u,s_a}_{i,j}} = \mathcal{CN}\left(c^{u,s_a}_{i,j} \mid \hat r^{c^{u,s_a}}_{i,j}, \tau^{r^{u,s_a}}_{i,j}\right),
\end{align}
where the mean $\hat r^{c^{u,s_a}}_{i,j}$ and variance $\tau^{c^{u,s_a}}_{i,j}$ are updated by GAMP (see lines 9-10 of Algorithm 1). Then, the message from $f^{c^{u,s_a}}_{i,j}$ to $ s^{u,s_a}_{i,j}$ can be derived as
\begin{align}
\label{yitaupdate}
	&\Delta^{f^{c^{u,s_a}}_{i,j}}_{s^{u,s_a}_{i,j}} 
	\propto
	\int_{c^{u,s_a}_{i,j}} p\left(c^{u,s_a}_{i, j} \mid s^{u,s_a}_{i, j}\right)  \Delta_{f^{c^{u,s_a}}_{i,j}}^{c^{u,s_a}_{i,j}}
	\nonumber \\
	&= \eta_{i,j}^{u,s_a} \delta\left(s_{i, j}^{u,s_a}-1\right) + (1-\eta_{i,j}^{u,s_a})\delta\left(s_{i, j}^{u,s_a}+1\right),
\end{align}
where $\eta_{i,j}^{u,s_a}=\frac{\eta_{i,j}^{\text{A}^{u,s_a}}}{\mathcal{CN}\left(0 \mid \hat r^{c^{u,s_a}}_{i,j}, \tau^{r^{u,s_a}}_{i,j}\right) +\eta_{i,j}^{\text{A}^{u,s_a}}}$, and $\eta_{i,j}^{\text{A}^{u,s_a}}=\sum_{k=1}^K \omega_k^{u,s_a} \mathcal{CN}(0 \mid \mu_k^{u,s_a}-\hat r^{c^{u,s_a}}_{i,j}, \phi_k^{u,s_a}+\tau^{r^{u,s_a}}_{i,j})$. Subsequently, MRF utilizes $\Delta^{f^{c^{u,s_a}}_{i,j}}_{s^{u,s_a}_{i,j}} $ to exploit the 2D block sparsity of the channel matrix, where the concrete message derivations are given in Appendix \ref{App2}. With the output message $\Delta_{f^{c^{u,s_a}}_{i,j}}^{s^{u,s_a}_{i,j}}$ from MRF, the refined prior of $c^{u,s_a}_{i,j}$ is given by
\begin{align}
	\label{TSEend}
	&\Delta^{f^{c^{u,s_a}}_{i,j}}_{c^{u,s_a}_{i,j}} 
	\propto
	\int_{s^{u,s_a}_{i,j}} p\left(c^{u,s_a}_{i,j} \mid s^{u,s_a}_{i, j}\right)  \Delta_{f^{c^{u,s_a}}_{i,j}}^{s^{u,s_a}_{i,j}} 
	= (1 - \zeta_{i,j}^{u,s_a}) \nonumber \\
	&\times \delta(c_{i,j}^{u,s_a}) + \zeta_{i,j}^{u,s_a} \sum_{k=1}^K \omega_k^{u,s_a} \mathcal{CN}\left(c_{i, j}^{u,s_a}\mid \mu^{u,s_a}_k, \phi^{u,s_a}_k\right),
\end{align}
where $\zeta_{i,j}^{u,s_a}$ is given in (\ref{zeta}). Then, combining the refined prior and the likelihood in (\ref{likec}), the posterior distribution of $c^{u,s_a}_{i,j}$ in this initial phase can be recognized as a BGM distribution, given as 
\begin{align}
	\label{cpI}
	\Delta_{c^{u,s_a}_{i,j}}^{\mathrm I} 
	\propto 
	&\Delta^{f^{c^{u,s_a}}_{i,j}}_{c^{u,s_a}_{i,j}}
	\Delta_{f^{c^{u,s_a}}_{i,j}}^{c^{u,s_a}_{i,j}}  
	=(1 - \chi_{i,j}^{u,s_a})\delta(c_{i,j}^{u,s_a})
	+ \chi_{i,j}^{u,s_a} \nonumber \\ &\times \sum_{k=1}^{K}\bar{\omega}_{i,j,k}^{u,s_a}\mathcal{CN}(c_{i,j}^{u,s_a}\mid \theta_{i,j,k}^{u,s_a}, \varphi_{i,j,k}^{u,s_a}),
\end{align}
where 
\begin{align}
\label{poststart}
&\varphi_{i,j,k}^{u,s_a}=
\left((\phi_k^{u,s_a})^{-1} + (\tau_{i,j}^{r^{u,s_a}})^{-1}\right)^{-1}, \\
&\theta_{i,j,k}^{u,s_a}=
\varphi_{i,j,k}^{u,s_a}
\left(\frac{\mu_k^{u,s_a}}{\phi_k^{u,s_a}} +
\frac{\hat{r}_{i,j}^{c^{u,s_a}}}{\tau_{i,j}^{r^{u,s_a}}}\right),\\
&\bar{\omega}_{i,j,k}^{u,s_a} = 
\frac{\omega_k^{u,s_a} 
	\mathcal{CN}\left(0 \mid \mu_k^{u,s_a}-\hat r^{c^{u,s_a}}_{i,j}, \phi_k^{u,s_a}+\tau^{r^{u,s_a}}_{i,j}\right)}{\eta_{i,j}^{\text{A}^{u,s_a}}},\label{omegaupdate} \\
&\chi_{i,j}^{u,s_a} = 
\frac{\zeta_{i,j}^{u,s_a}
	\eta_{i,j}^{\text{A}^{u,s_a}}
}{
	(1-\zeta_{i,j}^{u,s_a}) \mathcal{CN}(0\mid \hat{r}_{i,j}^{c^{u,s_a}}, \tau_{i,j}^{r^{u,s_a}})
	+\zeta_{i,j}^{u,s_a}
	\eta_{i,j}^{\text{A}^{u,s_a}}}.\label{Chiupdate}
\end{align}
Finally, the variance and mean of $c^{u,s_a}_{i,j}$ with respect to (\ref{cpI}) can be expressed as
\begin{align}
\label{cupdate}
	&\tau^{\mathrm I c^{u,s_a}}_{i,j} = \chi_{i,j}^{u,s_a} \sum_{k=1}^{K}\bar{\omega}_{i,j,k}^{u,s_a} (
	\left|\theta_{i,j,k}^{u,s_a}\right|^2 + \varphi_{i,j,k}^{u,s_a})
	-\left|\hat c^{\mathrm I^{u,s_a}}_{i,j}\right|^2, \nonumber \\
	&\hat c^{\mathrm I^{u,s_a}}_{i,j} = \chi_{i,j}^{u,s_a} \sum_{k=1}^{K}\bar{\omega}_{i,j,k}^{u,s_a} \theta_{i,j,k}^{u,s_a}.
\end{align}
Note that the previous derived algorithm needs the full knowledge of the noise variance and the parameters in prior BGM
i.e., $\bm{\bar{\theta}}_{s_a}\triangleq\{\sigma^2_{s_a},\omega_k^{u,s_a}, \mu_k^{u,s_a}, \phi_k^{u,s_a}, \forall k, u\}$. We now adopt the EM algorithm to learn these parameters, which is
an iterative technique that increases a lower bound on the likelihood at each iteration. Then, the update of $\bm{\bar{\theta}}_{s_a}$ in the $(t_{\mathrm{I}}+1)$-th iteration is given as
\begin{align}
\label{noiseVarianceUpdate}
	\bm{\bar{\theta}}^{t_{\mathrm{I}}+1}_{s_a} = \arg \max \limits_{\bm{\bar{\theta}}} 
	\mathrm{E}[\log p\left(\mathbf Z_{s_a}, \mathbf{C}_{s_a}, \mathbf Y_{s_a} \mid \bm{\bar{\theta}} \right) \mid \mathbf Y_{s_a}, \bm{\bar{\theta}}^{t_{\mathrm{I}}}_{s_a}],
\end{align}
where $\mathbf Z_{s_a}$ and $\mathbf{C}_{s_a}$ are treated as the hidden variables and the expectation is with respect to $p\left(\mathbf Z_{s_a},\mathbf{C}_{s_a} \mid \mathbf Y_{s_a},\bm{\bar{\theta}}^{t_{\mathrm{I}}}_{s_a}\right)$. With the previous approximated posterior distribution, we can derive the update of the hyperparameters as

\begin{align}
\label{noiseupdate}
	&(\sigma^2_{s_a})^{t_{\mathrm{I}}+1} = \frac{1}{M_{\mathrm p}N_{\mathrm p}N_zN_y} \sum_{m,j}\left|y_{m,j}^{s_a} - \hat z_{m,j}^{s_a}\right|^2 + \tau_{m,j}^{z^{s_a}},\\
	\label{muupdate}
	&(\mu_k^{u,s_a})^{t_{\mathrm{I}}+1} = \frac{\sum_{i,j}
		\chi_{i,j}^{u,s_a} \bar{\omega}_{i,j,k}^{u,s_a}
		\theta_{i,j,k}^{u,s_a}}{\sum_{i,j} \chi_{i,j}^{u,s_a} \bar{\omega}_{i,j,k}^{u,s_a}
	},\\
	&(\phi_k^{u,s_a})^{t_{\mathrm{I}}+1} = \frac{\sum_{i,j} \chi_{i,j}^{u,s_a} \bar{\omega}_{i,j,k}^{u,s_a}\left(
		\left|\theta_{i,j,k}^{u,s_a}-\mu_k^{u,s_a}\right|^2 + \varphi_{i,j,k}^{u,s_a} \right)}{\sum_{i,j} \chi_{i,j}^{u,s_a}  \bar{\omega}_{i,j,k}^{u,s_a}
	},\\
	&(\omega_{k}^{u,s_a})^{t_{\mathrm{I}}+1} = \frac{\sum_{i,j} \chi_{i,j}^{u,s_a}  \bar{\omega}_{i,j,k}^{u,s_a}}{\sum_{i,j} \chi_{i,j}^{u,s_a}}.\label{EMend}
\end{align}
\vspace{.3cm}

Building upon the message expressions and EM update rules, we propose the MRF-BGM-AMP algorithm for initial channel estimation, and we denote $x_{m,i}^{\omega^{u}}$ as the $(m,u(Q_{\mathrm I}+1)(L+1)+i)$-th element of $\mathbf X^{\omega}$ as shown in Algorithm 1.  The lines 3-10 represent the GAMP, lines 11-18 are expressions derived using message passing rules, and the line 21 is the EM update. Proper initialization of the
hyperparameters is crucial for EM update which refers to \cite{c25}, and the damping factor is leveraged in the GAMP part to help the convergence of the algorithm \cite{sss}. 

\begin{algorithm}[hbt!]
	\footnotesize
	\caption{MRF-BGM-AMP}\label{MRF-BGM-AMP}
	\begin{algorithmic}[1]
		\REQUIRE{Recieved signals $\mathbf Y_{s_a}$, sensing matrix $\mathbf X^{\omega}$; the maximum number of iterations $T_{\text I}$ and $T_{\text{mrf}}$, and the termination threshold $\eta_{\text I}$}
		\ENSURE The estimated BEM coefficient $\hat c^{\mathrm I^{u,s_a}}_{i,j}$, variance $\tau^{\mathrm I c^{u,s_a}}_{i,j}$ and hyperparameters $\bm{\bar{\theta}}_{s_a}$.
		\STATE \textbf{Initialization}: choose $\bm{\bar{\theta}}_{s_a}$; $\forall i,j,u,s_a$: choose $\hat c^{\mathrm I^{u,s_a}}_{i,j}, \tau^{\mathrm I c^{u,s_a}}_{i,j}$, $\xi^{\text{L}^{u,s_a}}_{i,j}=\xi^{\text{R}^{u,s_a}}_{i,j}=\xi^{\text{T}^{u,s_a}}_{i,j}=\xi^{\text{B}^{u,s_a}}_{i,j}=0.5$;
		$\forall m,j,s_a$: $\hat s^{z^{s_a}}_{m,j}=0$;
		\FOR{$t_{\mathrm{I}}=1$ to $T_{\text{I}}$}
		\STATE $\forall m,j,s_a$: $\tau_{m,j}^{p^{s_a}}=\sum_{u,i}\left|x_{m,i}^{\omega^{u}}\right|^2 \tau^{\mathrm I c^{u,s_a}}_{i,j}$
		\STATE $\forall m,j,s_a$: $\hat{p}_{m,j}^{s_a}=\sum_{u,i} x_{m,i}^{\omega^u} \hat c^{\mathrm I^{u,s_a}}_{i,j}-\tau_{m,j}^{p^{s_a}} \hat s^{z^{s_a}}_{m,j}$.
		\STATE $\forall m,j,s_a$: $\tau_{m,j}^{z^{s_a}}=\tau_{m,j}^{p^{s_a}}\sigma^2_{s_a} / \left(\tau_{m,j}^{p^{s_a}}+\sigma^2_{s_a}\right)$ 
		\STATE $\forall m,j,s_a$: $\hat z_{m,j}^{s_a}=\left(\tau_{m,j}^{p^{s_a}}y_{m, j}^{s_a}+\sigma^2_{s_a} \hat{p}_{m,j}^{s_a}\right)/\left(\tau_{m,j}^{p^{s_a}}+\sigma^2_{s_a}\right)
		$ 
		\STATE $\forall m, j, s_a$: $\tau_{m,j}^{s^{s_a}}=(\tau_{m,j}^{p^{s_a}}-\tau_{m,j}^{z^{s_a}}) /(\tau_{m,j}^{p^{s_a}})^2$
		\STATE $\forall m, j, s_a$: $\hat s^{z^{s_a}}_{m,j}=(\hat z_{m,j}^{s_a}-\hat{p}_{m,j}^{s_a}) / \tau_{m,j}^{p^{s_a}}$
		\STATE $\forall i, j, u, s_a$: $\tau^{r^{u,s_a}}_{i,j}=(\sum_m\left|x_{m,i}^{\omega^u}\right|^2 \tau_{m,j}^{s^{s_a}})^{-1}$
		\STATE $\forall i, j, u, s_a$: $r^{c^{u,s_a}}_{i,j}=\hat c^{\mathrm I^{u,s_a}}_{i,j}+\tau^{r^{u,s_a}}_{i,j} \sum_m x_{m,i}^{\omega^{u^*}} \hat s^{z^{s_a}}_{m,j}$
		\STATE $\forall i, j, u, s_a$: Update $\eta_{i,j}^{u,s_a}$ via (\ref{yitaupdate})
		\STATE $\backslash \backslash$ \textbf{MRF Module}.
		\FOR{$t_{\text{mrf}}=1$ to $T_{\text{mrf}}$}
		\STATE $\forall i,j,u,s_a$: Update $\xi^{\text{L}^{u,s_a}}_{i,j},\xi^{\text{R}^{u,s_a}}_{i,j},\xi^{\text{T}^{u,s_a}}_{i,j},\xi^{\text{B}^{u,s_a}}_{i,j}$ via (\ref{xiupdate})
		\ENDFOR
		\STATE $\forall i,j,u,s_a$: Update $\zeta_{i,j}^{u,s_a}$ and $\chi_{i,j}^{u,s_a}$ via (\ref{zitaupdate}) and (\ref{Chiupdate})
		\STATE $\forall i,j,k,u,s_a$: Update $\varphi_{i,j,k}^{u,s_a}$, $\theta_{i,j,k}^{u,s_a}$, and $\bar{\omega}_{i,j,k}^{u,s_a}$ via (\ref{poststart})-(\ref{omegaupdate})
		\STATE $\forall i,j,u,s_a$: Update $\hat c^{\mathrm I^{u,s_a}}_{i,j}(t_{\mathrm{I}}+1)$ and $\tau^{\mathrm I c^{u,s_a}}_{i,j}(t_{\mathrm{I}}+1)$ via (\ref{cupdate})
		\STATE \textbf{if} $\left\|\hat{\mathbf{ C}}_{s_a}(t_{\mathrm{I}}+1)-\hat{\mathbf{ C}}_{s_a}\right\|^2_{\mathrm{F}} \leq \eta_{\text I} \left\|\hat{\mathbf{ C}}_{s_a}\right\|^2_{\mathrm{F}}$: break \textbf{end if}
		\STATE $\backslash \backslash$ \textbf{EM Update}
		\STATE $\forall s_a$: Update $\bm{\bar{\theta}}_{s_a}$ via (\ref{noiseupdate})-(\ref{EMend})
		\ENDFOR
	\end{algorithmic}
\end{algorithm}
\section{Enhancement via Cooperative Satellite Constellations}
\label{IV}
In this section, we detail our algorithm for joint device identification, channel estimation, and symbol detection, enhanced by cooperative LEO satellite constellations. We explore two cooperative modes: centralized and distributed modes. Firstly, we describe the method for identifying active devices, and then the algorithm for initial symbol detection. Finally, we introduce the joint refinement process for channel estimation and symbol detection.
\vspace{-0.4cm}
\subsection{Centralized Cooperative Mode}
In the centralized mode, all the computation is performed centrally at the gateway or one specific satellite with high computation power, which requires each edge satellite to feed back their own received signal to the central server for further processing.
\subsubsection{Device identification}
To identify active devices, the Algorithm 1 is firstly adopted to get the initial channel estimation $\hat c^{\mathrm I^{u,s_a}}_{i,j}$. Then, by aggregating the estimated channel energy at the central node, the activity indicator of the $u$-th device is given by
\begin{align}
\label{DI}
	\hat \lambda_u = \mathbb{I}\left\{\frac{1}{S_a}\sum_{s_a,i,j}\left|\hat c^{\mathrm I^{u,s_a}}_{i,j}\right|^2 > \eta_{\lambda}\right\},
\end{align}
where $\mathbb{I}\{\cdot\}$ is the indicator function,  $\eta_{\lambda}$ is an empirical predefined threshold, and we set $\eta_{\lambda}=10^{-2}$ in the simulations. Then, the set of estimated active devices is given by $\hat{\mathcal{U}}_a=\{u\mid \hat \lambda_u=1, u=0,\dots,U-1\}$. 
\subsubsection{Initial symbol detection}
According to (\ref{modulationstart})-(\ref{YDD}), the received signal at the $n_a$-th antenna of the $s_a$-th satellite can be written as
\begin{align}
	\mathbf{y}^{\text{DD}}_{n_a,s_a}=\sum_{u \in \mathcal{U}_a} \bar{\mathbf{H}}_{u,s_a}^{n_a} \mathbf{x}_{u} +  \mathbf{e}_{s_a}^{n_a},
\end{align}
where $\bar{\mathbf{H}}_{u,s_a}^{n_a}=\left(\mathbf{F}_N \otimes \mathbf{I}_M\right) {\mathbf{H}}_{u,s_a}^{n_a}\left(\mathbf{F}_N^{\mathrm{H}} \otimes \mathbf{I}_M\right)$, $\mathbf{H}_{u,s_a}^{n_a} \in \mathbb{C}^{MN\times MN}$ with the $(n,m)$-th element $h_{u,s_a}^{n_a}[n+M_{\text{cp}}, (n-m)_{MN}]$, and $\mathbf{x}_{u} = \operatorname{vec}(\mathbf{X}^{\text{DD}}_{u})$. At the central node, the received signals of all the satellites are collected in $\mathbf{y}^{\text{DD}}=[\mathbf{y}^{\text{DD}^{\mathrm{T}}}_{0,0},\dots,\mathbf{y}^{\text{DD}^{\mathrm{T}}}_{n_a,s_a},\dots,\mathbf{y}^{\text{DD}^{\mathrm{T}}}_{N_zN_y-1,S_a-1}]^{\mathrm{T}}$, and then we have the following relationship
\begin{align}
\label{intialSD}
	\mathbf{y}^{\text{DD}}=\bar{\mathbf{H}} \mathbf{x} +  \mathbf{e},
\end{align}
where $\bar{\mathbf{H}}=[\bar{\mathbf{H}}_{0,0}^{\mathrm{T}},\dots,\bar{\mathbf{H}}_{n_a,s_a}^{\mathrm{T}},\dots,\bar{\mathbf{H}}_{N_zN_y-1,S_a-1}^{\mathrm{T}}]^{\mathrm{T}}$, $\bar{\mathbf{H}}_{n_a,s_a}=[\bar{\mathbf{H}}_{0,s_a}^{n_a},\dots,\bar{\mathbf{H}}_{u,s_a}^{n_a},\dots,\bar{\mathbf{H}}_{U-1,s_a}^{n_a}]$ for $u\in\mathcal{U}_a$, $\mathbf{x}=[\mathbf{x}^{\mathrm{T}}_{0},\dots,\mathbf{x}^{{\mathrm{T}}}_{u}\dots,\mathbf{x}^{{\mathrm{T}}}_{U-1}]^{\mathrm{T}}$ for $u\in\mathcal{U}_a$, and $\mathbf{e}=[\mathbf{e}_{0}^{0^{\mathrm{T}}},\dots,\mathbf{e}_{S_a-1}^{{N_zN_y-1}^{\mathrm{T}}}]^{\mathrm{T}}$. Note that $\mathcal{U}_a$ can be estimated by $\hat{\mathcal{U}}_a$, and the channel coefficient $\hat h_{u,s_a}^{n_a}[n, l^{\prime}]$ can be reconstructed by $\hat c^{\mathrm I^{u,s_a}}_{i,j}$ according to (\ref{has}). Then, the GAMP algorithm with uniform prior \cite{sss} is utilized for symbol detection. $\hat{\mathbf{x}}_u^{\mathrm I}$ including both pilots and estimated data symbols and the corresponding variance $\bm{\tau}_u^{x_{\mathrm I}}$ of active devices can be obtained.
\subsubsection{Jointly refine channel estimation and symbol detection} 
We adopt a higher resolution $R=2$ with the larger order of BEM $Q_{\mathrm R} \geq 2\left\lceil 2 N \bar{\nu}_{\max}\right\rceil$ to improve the modeling accuracy for TSLs. Here, the received signals in the time-angle domain are utilized to jointly refine channel estimation and symbol detection for active devices, where the only difference from (\ref{Ymat}) is that there is no demodulation matrix $\mathbf{F}_N \otimes \mathbf{I}_M$, i.e.,
\begin{align}
\label{YmatT}
	\mathbf{y}_{n_a,s_a}^{\prime}=\sum_{u\in\mathcal{U}_a}\sum_{q=0}^{Q_{\mathrm R}}
	&\operatorname{diag}\left(\mathbf{b}_q\right) 
	\mathbf{F}_{MN}^{\mathrm{H}}  
	\operatorname{diag}\left(\mathbf{F}_{1:L+1} \mathbf{c}_{q,u,s_a}^{n_a}\right) \mathbf{F}_{MN} \nonumber \\
	&\times \left(\mathbf{F}_N^{\mathrm{H}} \otimes \mathbf{I}_M\right) 
	\mathbf{x}_u + \mathbf{e}_{n_a,s_a}^{\prime},
\end{align}
where $\mathbf{e}_{n_a,s_a}^{\prime}$ represents the noise in the time-angle domain.
To facilitate the algorithm design, we introduce the auxiliary variables $\mathbf x^{\text F}_{u}=\mathbf{F}_{\text b} \mathbf{x}_u$,
$\mathbf c^{\text F}_{q,u,n_a, s_a} = \mathbf{F}_{1:L+1} \mathbf{c}_{q,u,s_a}^{n_a}$, 
$\mathbf d^{\text{P}}_{q, u, n_a, s_a}
=\operatorname{diag}\left(\mathbf c^{\text F}_{q,u,n_a, s_a}\right) \mathbf x^{\text F}_{u}$, 
$\mathbf d^{\text{W}}_{q, n_a, s_a}=\sum_u \mathbf d^{\text{P}}_{q, u, n_a, s_a}$,
$\mathbf d^{\text F}_{q, n_a, s_a}=\mathbf{F}_{MN}^{\mathrm{H}} \mathbf d^{\text{W}}_{q, n_a, s_a}$,
$\mathbf d^{\text{B}}_{q, n_a, s_a}
=\operatorname{diag}\left(\mathbf{b}_q\right) \mathbf d^{\text F}_{q, n_a, s_a}$, and
$\mathbf g_{n_a, s_a}=\sum_q \mathbf d^{\text{B}}_{q, n_a, s_a}$, 
where $\mathbf{F}_{\text b}\triangleq\mathbf{F}_{MN}
\left(\mathbf{F}_N^{\mathrm{H}} \otimes \mathbf{I}_M\right)$. Then, the posterior distribution of the variables is given by
\begin{align}
\label{jointpost}
	&p\left( {{\bf{x}},{\bf{c}},{{\bf{x}}^{\rm{F}}},{{\bf{c}}^{\rm{F}}},{{\bf{d}}^{\rm{P}}},{{\bf{d}}^{\rm{W}}},{{\bf{d}}^{\rm{F}}},{{\bf{d}}^{\rm{B}}},{\bf{g}}\mid {\bf{y}}} \right) \propto \nonumber \\
	&\prod\limits_{{s_a},{n_a}} {\{ p} ({\bf{y}}_{{n_a},{s_a}}^\prime \mid {{\bf{g}}_{{n_a},{s_a}}})p({{\bf{g}}_{{n_a},{s_a}}}\mid {\bf{d}}_{0,{n_a},{s_a}}^{\rm{B}}, \cdots ,{\bf{d}}_{Q,{n_a},{s_a}}^{\rm{B}})\nonumber \\
	&\prod\limits_q {[p} ({\bf{d}}_{q,{n_a},{s_a}}^{\rm{B}}\mid {\bf{d}}_{q,{n_a},{s_a}}^{\rm{F}})p({\bf{d}}_{q,{n_a},{s_a}}^{\rm{F}}\mid {\bf{d}}_{q,{n_a},{s_a}}^{\rm{W}}) \nonumber \\
	&p({\bf{d}}_{q,{n_a},{s_a}}^{\rm{W}}\mid {\bf{d}}_{q,0,{n_a},{s_a}}^{\rm{P}}, \cdots ,{\bf{d}}_{q,U - 1,{n_a},{s_a}}^{\rm{P}})\nonumber \\
	&\prod\limits_u p ({\bf{d}}_{q,u,{n_a},{s_a}}^{\rm{P}}\mid {\bf{c}}_{q,u,{n_a},{s_a}}^{\rm{F}},{\bf{x}}_u^{\rm{F}})p({\bf{c}}_{q,u,{n_a},{s_a}}^{\rm{F}}\mid {\bf{c}}_{q,u,{s_a}}^{{n_a}}) \nonumber \\
	&p({\bf{c}}_{q,u,{s_a}}^{{n_a}})]\} 
	\prod\limits_u {p({\bf{x}}_u^{\rm{F}}\mid {{\bf{x}}_u})p({{\bf{x}}_u})},
\end{align}
where ${\bf{x}}=\{{\bf{x}}_u\mid \forall u\}$, ${\bf{c}}=\{{\bf{c}}_{q,u,{s_a}}^{{n_a}}\mid \forall q,u,n_a,s_a\}$, 
${{\bf{x}}^{\rm{F}}}=\{{\bf{x}}_u^{\rm{F}}\mid \forall u\}$,
${{\bf{c}}^{\rm{F}}}=\{{\bf{c}}_{q,u,{n_a},{s_a}}^{\rm{F}}\mid \forall q,u,n_a,s_a \}$,
${{\bf{d}}^{\rm{P}}}=\{\mathbf d^{\text{P}}_{q, u, n_a, s_a}\mid \forall q,u,n_a,s_a\}$,
${{\bf{d}}^{\rm{W}}}=\{\mathbf d^{\text{W}}_{q, n_a, s_a}\mid \forall q,n_a,s_a\}$,
${{\bf{d}}^{\rm{F}}}=\{\mathbf d^{\text F}_{q, n_a, s_a}\mid \forall q,n_a,s_a\}$,
${{\bf{d}}^{\rm{B}}}=\{\mathbf d^{\text{B}}_{q, n_a, s_a}\mid \forall q,n_a,s_a\}$,
${\bf{g}}=\{\mathbf g_{n_a, s_a}\mid \forall n_a,s_a\}$, and 
${\bf{y}}=\{\mathbf{y}_{n_a,s_a}^{\prime}\mid \forall n_a,s_a\}$.

\begin{table*}
	\scriptsize
	\caption{Notations of Factor Nodes} \label{factorTable}
	\vspace{-.4cm}
	\begin{center}
		\begin{tabular}{|c|c|c|}
			\hline Factor & Distribution & Function\\
			\hline $f^{\mathbf y}_{n_a,s_a}$ & $p(\mathbf{y}_{n_a,s_a}^{\prime} \mid \mathbf g_{n_a, s_a})$ & $\mathcal{CN}(\mathbf y_{n_a,s_a}^{\prime}\mid \mathbf g_{n_a, s_a}, \sigma^2_{s_a}\mathbf I_{MN})$\\
			\hline $f^{\mathbf g}_{n_a,s_a}$ & $p(\mathbf g_{n_a, s_a} \mid \mathbf d^{\text{B}}_{0, n_a, s_a}, \cdots, \mathbf d^{\text{B}}_{Q, n_a, s_a}) $ & $\delta(\mathbf g_{n_a, s_a}-\sum_q \mathbf d^{\text{B}}_{q, n_a, s_a})$\\
			\hline $f^{\text B}_{q,n_a,s_a}$ & $p(\mathbf d^{\text{B}}_{q, n_a, s_a} \mid \mathbf d^{\text F}_{q, n_a, s_a})$ & $\delta(\mathbf d^{\text{B}}_{q, n_a, s_a}
			-\operatorname{diag}\left(\mathbf{b}_q\right) \mathbf d^{\text F}_{q, n_a, s_a})$\\
			\hline $f^{\mathbf d_{\text F}}_{q,n_a,s_a}$  & $p(\mathbf d^{\text F}_{q, n_a, s_a} \mid \mathbf d^{\text{W}}_{q, n_a, s_a})$ & $\delta(\mathbf d^{\text F}_{q, n_a, s_a}-\mathbf{F}_{MN}^{\mathrm{H}} \mathbf d^{\text{W}}_{q, n_a, s_a})$\\
			\hline $f^{\text W}_{q,n_a,s_a}$ & $p(\mathbf d^{\text{W}}_{q, n_a, s_a} \mid \mathbf d^{\text{P}}_{q,0, n_a, s_a}, \cdots, \mathbf d^{\text{P}}_{q, U-1, n_a, s_a})$ & $\delta(\mathbf d^{\text{W}}_{q, n_a, s_a}-\sum_u \mathbf d^{\text{P}}_{q, u, n_a, s_a})$\\
			\hline $f^{\text{P}}_{q, u, n_a, s_a}$ & $p(\mathbf d^{\text{P}}_{q,u, n_a, s_a} \mid \mathbf c^{\text F}_{q,u, n_a, s_a}, \mathbf x^{\text F}_{u})$ & $\delta(\mathbf d^{\text{P}}_{q, u, n_a, s_a}
			-\operatorname{diag}\left(\mathbf c^{\text F}_{q,u,n_a, s_a}\right) \mathbf x^{\text F}_{u})$\\
			\hline $f^{\mathbf c_{\text F}}_{q, u, n_a, s_a}$ & $p(\mathbf c^{\text F}_{q,u,n_a, s_a} \mid \mathbf{c}_{q,u,s_a}^{n_a})$ & $\delta(\mathbf c^{\text F}_{q,u,n_a, s_a} - \mathbf{F}_{1:L+1} \mathbf{c}_{q,u,s_a}^{n_a})$\\
			\hline $f^{\mathbf x_{\text F}}_{u}$ & $p({\bf{x}}_u^{\rm{F}}\mid {{\bf{x}}_u})$& $\delta(\mathbf x^{\text F}_{u}-\mathbf{F}_{\text b} \mathbf{x}_u)$\\
			\hline
		\end{tabular}
		\label{factors}
	\end{center}
\end{table*}
\begin{figure}[!htb]
	\centering
	\includegraphics[width=3.4in]{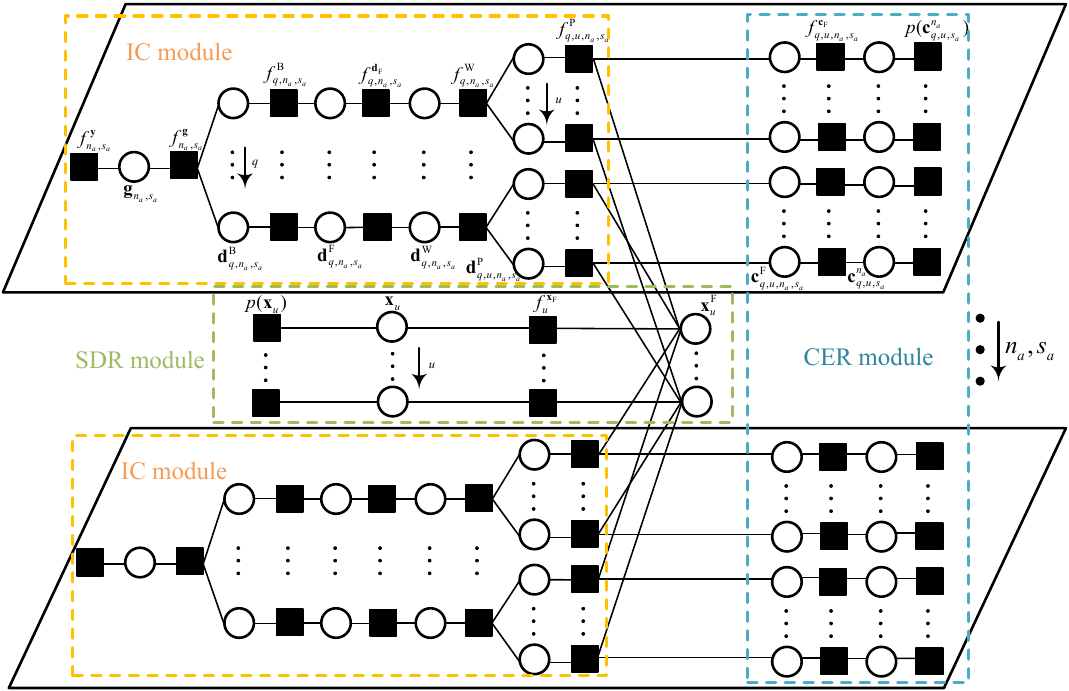}
	\caption{Factor graph representation for refinement scheme. }
	\label{FactorGraph}
\end{figure}
The factor graph representation of (\ref{jointpost}) is shown in Fig. \ref{FactorGraph}, and the corresponding factors are summarized in Table \ref{factorTable}. For sake of low computations, we assign the non-informative prior for $\mathbf{c}_{q,u,s_a}^{n_a}$ and $\mathbf x_{u}$, i.e., $p(\mathbf{c}_{q,u,s_a}^{n_a})=\mathcal{CN}(\mathbf{c}_{q,u,s_a}^{n_a}\mid \mathbf 0_{L+1}, \infty \mathbf I_{L+1})$ and $p(\mathbf x_{u})=\mathcal{CN}(\mathbf x_{u}\mid\mathbf 0_{MN}, \infty \mathbf I_{MN})$, which is enough for achieving superior performance as shown in the simulations. Since exact BP rule is difficult to implement in Fig. \ref{FactorGraph}, we resort to EP which extends BP by enforcing an exponential family constraint on the messages. Based on the EP and central limit theorem (CLT), we aims to approximate the posterior distribution $p(\mathbf x_u \mid \mathbf y)$ and $p(\mathbf{c}_{q,u,s_a}^{n_a}\mid \mathbf y)$, leading to the MAP estimations $\hat{\mathbf{x}}_u$ and $\hat{\mathbf{c}}_{q,u,s_a}^{n_a}$. 
As shown in Fig. \ref{FactorGraph}, we divide the whole receiver structure into three modules: interference cancellation (IC) module addresses the linear signal models to mitigate both inter-user and inter-component interference within the BEM model, subsequently providing likelihoods for $\mathbf c^{\text F}_{q,u,n_a, s_a}$ and $\mathbf x^{\text F}_{u}$; The channel estimation refinement (CER) and symbol detection refinement (SDR) modules further enhance channel estimation and symbol detection, respectively, utilizing soft information from the IC module. 
For EP, the projection
of one density onto the Gaussian family is a key concept, denoted as
\begin{align}
	\operatorname{Proj}[p^{\prime}(x)] &\triangleq \arg \min _{p \in \mathcal{G}} \mathrm{KL}[p^{\prime}(x) \| p(x)] \nonumber \\
	&=\mathcal{CN}(x \mid \mu, \sigma^2),
\end{align}
where $\mathrm{KL}[\cdot]$ represents the KL-divergence, $\mathcal{G}$ is the family of Gaussian densities, $\mu =\int x p^{\prime}(x) \mathrm{d} x$, and $\sigma^2=\int|x-\mu|^2 p^{\prime}(x) \mathrm{d} x$. 

In Appendix \ref{App3}, we derive the messages in the IC module. It provides the likelihoods for CER module and SDR module, respectively, which are given as
\begin{align}
	\label{fPcF}
	&\Delta^{f^{\text{P}}_{q, u, n_a, s_a}}_{\mathbf c_{q,u,n_a, s_a}^{\text F}}=\mathcal{CN}(\mathbf c^{\text F}_{q,u,n_a, s_a}\mid \overrightarrow{\mathbf{c}}^{\text F}_{q,u,n_a, s_a}, \operatorname{diag}(\overrightarrow{\bm{\tau}}^{c_{\text F}}_{q,u,n_a, s_a}))\\
	&\Delta^{f^{\text{P}}_{q, u, n_a, s_a}}_{\mathbf x^{\text F}_{u}}=\mathcal{CN}(\mathbf x^{\text F}_{u}\mid \overleftarrow{\mathbf{x}}^{\text F}_{q, u, n_a, s_a}, \operatorname{diag}(\overleftarrow{\bm{\tau}}^{x_{\text F}}_{q, u, n_a, s_a})).\label{fPxF}
\end{align}
With the likelihood in (\ref{fPcF})  and $\Delta^{\mathbf{c}_{q,u,s_a}^{n_a}}_{f^{\mathbf c_{\text F}}_{q, u, n_a, s_a}}=\mathcal{CN}(\mathbf{c}_{q,u,s_a}^{n_a} \mid
\overleftarrow{\mathbf{c}}_{q,u,s_a}^{n_a}, 
\overleftarrow{\tau}_{q,u,s_a}^{c_{n_a}}\mathbf I_{L+1})$ passing to CER module, the posterior distribution of $\mathbf{c}_{q,u,s_a}^{n_a}$ can be obtained as
\begin{align}
\label{postcr}
	\Delta_{\mathbf{c}_{q,u,s_a}^{n_a}} 
	&\propto
	\operatorname{Proj}\left[\Delta^{\mathbf{c}_{q,u,s_a}^{n_a}}_{f^{\mathbf c_{\text F}}_{q, u, n_a, s_a}}
	\int_{\mathbf c^{\text F}_{q,u,n_a, s_a}} 
	f^{\mathbf c_{\text F}}_{q, u, n_a, s_a}
	\Delta^{f^{\text{P}}_{q, u, n_a, s_a}}_{\mathbf c_{q,u,n_a, s_a}^{\text F}}\right] \nonumber \\
	&\propto
	\mathcal{CN}(\mathbf{c}_{q,u,s_a}^{n_a} \mid \hat{\mathbf{c}}_{q,u,s_a}^{n_a}, \tau_{q,u,s_a}^{c_{n_a}} \mathbf I_{L+1}),
\end{align}
where
\begin{align}
	\label{overleftarrow_tau}
	&\overleftarrow{\tau}_{q,u,s_a}^{c_{n_a}}=\frac{1}{L+1}\operatorname{Tr}[\bm{\Sigma}_{q,u,s_a}^{c_{n_a}}],\\
	&\bm{\Sigma}_{q,u,s_a}^{c_{n_a}} =\nonumber \\ &(\mathbf{F}_{1:L+1}^{\mathrm{H}}\operatorname{diag}^{-1}(\overrightarrow{\bm{\tau}}^{c_{\text F}}_{q,u,n_a,s_a})\mathbf{F}_{1:L+1}+(\overleftarrow{\tau}_{q,u,s_a}^{c_{n_a}})^{-1}\mathbf I_{L+1})^{-1},\\
	&\hat{\mathbf{c}}_{q,u,s_a}^{n_a} = \bm{\Sigma}_{q,u,s_a}^{c_{n_a}}  
	(\mathbf{F}_{1:L+1}^{\mathrm{H}}\operatorname{diag}^{-1}(\overrightarrow{\bm{\tau}}^{c_{\text F}}_{q,u,n_a,s_a})\overrightarrow{\mathbf{c}}^{\text F}_{q,u,n_a, s_a} \nonumber \\
	&+\overleftarrow{\mathbf{c}}_{q,u,s_a}^{n_a}/\overleftarrow{\tau}_{q,u,s_a}^{c_{n_a}}).\label{LMMSEC}
\end{align}
Note that (\ref{LMMSEC}) can be recognized as the LMMSE estimate of a random vector $\mathbf{c}_{q,u,s_a}^{n_a}$ under likelihood $\mathcal{CN}(\overrightarrow{\mathbf{c}}^{\text F}_{q,u,n_a, s_a}\mid \mathbf{F}_{1:L+1} \mathbf{c}_{q,u,s_a}^{n_a}, \operatorname{diag}(\overrightarrow{\bm{\tau}}^{c_{\text F}}_{q,u,n_a, s_a}))$ and prior $\mathbf{c}_{q,u,s_a}^{n_a}\sim \mathcal{CN}(\mathbf{c}_{q,u,s_a}^{n_a} \mid
\overleftarrow{\mathbf{c}}_{q,u,s_a}^{n_a}, 
\overleftarrow{\tau}_{q,u,s_a}^{c_{n_a}}\mathbf I_{L+1})$. Then, the message $\Delta_{\mathbf{c}_{q,u,s_a}^{n_a}}^{f^{\mathbf c_{\text F}}_{q, u, n_a, s_a}}$ is computed as
\begin{align}
\label{ctocF}
	\Delta_{\mathbf{c}_{q,u,s_a}^{n_a}}^{f^{\mathbf c_{\text F}}_{q, u, n_a, s_a}}
	&\propto 
	\frac{\Delta_{\mathbf{c}_{q,u,s_a}^{n_a}}}{\Delta^{\mathbf{c}_{q,u,s_a}^{n_a}}_{f^{\mathbf c_{\text F}}_{q, u, n_a, s_a}}} \nonumber \\
	&\propto \mathcal{CN}(\mathbf{c}_{q,u,s_a}^{n_a} \mid
	\overrightarrow{\mathbf{c}}_{q,u,s_a}^{n_a}, 
	\overrightarrow{\tau}_{q,u,s_a}^{c^{n_a}}\mathbf I_{L+1}),
\end{align}
where $\overrightarrow{\tau}_{q,u,s_a}^{c_{n_a}}=((\tau_{q,u,s_a}^{c_{n_a}})^{-1}-(\overleftarrow{\tau}_{q,u,s_a}^{c_{n_a}})^{-1})^{-1}$ and $\overrightarrow{\mathbf{c}}_{q,u,s_a}^{n_a}=\overrightarrow{\tau}_{q,u,s_a}^{c_{n_a}}\left(\hat{\mathbf{c}}_{q,u,s_a}^{n_a} / \tau_{q,u,s_a}^{c_{n_a}}-\overleftarrow{\mathbf{c}}_{q,u,s_a}^{n_a}/
\overleftarrow{\tau}_{q,u,s_a}^{c_{n_a}}\right)$. Since non-informative prior is adopted, we update $\Delta^{\mathbf{c}_{q,u,s_a}^{n_a}}_{f^{\mathbf c_{\text F}}_{q, u, n_a, s_a}}$ as  $\Delta^{\mathbf{c}_{q,u,s_a}^{n_a}}_{f^{\mathbf c_{\text F}}_{q, u, n_a, s_a}}=\Delta_{\mathbf{c}_{q,u,s_a}^{n_a}}^{f^{\mathbf c_{\text F}}_{q, u, n_a, s_a}}$. Then, the message $\Delta_{\mathbf c_{q,u,n_a, s_a}^{\text F}}^{f^{\mathbf c_{\text F}}_{q, u, n_a, s_a}}$ is obtained as $\Delta_{\mathbf c_{q,u,n_a, s_a}^{\text F}}^{f^{\mathbf c_{\text F}}_{q, u, n_a, s_a}}=\mathcal{CN}(\mathbf c^{\text F}_{q,u,n_a, s_a}\mid \overleftarrow{\mathbf{c}}^{\text F}_{q,u,n_a, s_a}, \overleftarrow{{\tau}}^{c_{\text F}}_{q,u,n_a, s_a} \mathbf I_{MN})$ where $\overleftarrow{\mathbf{c}}^{\text F}_{q,u,n_a, s_a}=\mathbf{F}_{1:L+1} \overleftarrow{\mathbf{c}}_{q,u,s_a}^{n_a}$ and $\overleftarrow{{\tau}}^{c_{\text F}}_{q,u,n_a, s_a}=\overleftarrow{\tau}_{q,u,s_a}^{c_{n_a}}$. Now, we can update the posterior distribution of $\mathbf c^{\text F}_{q,u,n_a, s_a}$ as 
\begin{align}
\label{DeltacF}
	&\Delta_{\mathbf c_{q,u,n_a, s_a}}^{{\text F}} 
	\propto \Delta^{f^{\text{P}}_{q, u, n_a, s_a}}_{\mathbf c_{q,u,n_a, s_a}^{\text F}} \Delta_{\mathbf c_{q,u,n_a, s_a}^{\text F}}^{f^{\mathbf c_{\text F}}_{q, u, n_a, s_a}}
	\nonumber \\
	&\propto 
	\mathcal{CN}(\mathbf c^{\text F}_{q,u,n_a, s_a}\mid \hat{\mathbf{c}}^{\text F}_{q,u,n_a, s_a}, \operatorname{diag}(\bm{\tau}^{c_{\text F}}_{q,u,n_a, s_a})),
\end{align}
where
\begin{align}
	&\bm{\tau}^{c_{\text F}}_{q,u,n_a, s_a}=1\oslash((\overleftarrow{{\tau}}^{c_{\text F}}_{q,u,n_a, s_a})^{-1}\mathbf 1_{MN}+1\oslash \overrightarrow{\bm{\tau}}^{c_{\text F}}_{q,u,n_a, s_a}),\\
	&\hat{\mathbf{c}}^{\text F}_{q,u,n_a, s_a}=\bm{\tau}^{c_{\text F}}_{q,u,n_a, s_a}\odot(\overleftarrow{\mathbf{c}}^{\text F}_{q,u,n_a, s_a}/ \overleftarrow{{\tau}}^{c_{\text F}}_{q,u,n_a, s_a}
	\nonumber \\
	&\qquad \qquad \qquad \qquad \qquad+ \overrightarrow{\mathbf{c}}^{\text F}_{q,u,n_a, s_a}\oslash \overrightarrow{\bm{\tau}}^{c_{\text F}}_{q,u,n_a, s_a}).
\end{align}

In the SDR module, by combining the messages from $f^{\text{P}}_{q, u, n_a, s_a}$ to $\mathbf x^{\text F}_{u}$, we get the effective likelihood of $\mathbf x^{\text F}_{u}$ as
\begin{align}
\label{effecxF}
	\Delta_{f^{\mathbf x_{\text F}}_{u}
	}^{\mathbf x^{\text F}_{u}} 
	&\propto
	\prod_{q,n_a,s_a}\Delta^{f^{\text{P}}_{q, u, n_a, s_a}}_{\mathbf x^{\text F}_{u}} \nonumber \\
	&= \mathcal{CN}(\mathbf x^{\text F}_{u}\mid \overleftarrow{\bar{\mathbf{x}}}^{\text F}_{u}, \operatorname{diag}(\overleftarrow{\bar{\bm{\tau}}}^{x_{\text F}}_{u})),
\end{align}
where $\overleftarrow{\bar{\bm{\tau}}}^{x_{\text F}}_{u}=1\oslash \sum_{q,n_a,s_a} (1\oslash \overleftarrow{\bm{\tau}}^{x_{\text F}}_{q, u, n_a, s_a})$ and $\overleftarrow{\bar{\mathbf{x}}}^{\text F}_{u}=\overleftarrow{\bar{\bm{\tau}}}^{x_{\text F}}_{u}\odot\sum_{q,n_a,s_a} (\overleftarrow{\mathbf{x}}^{\text F}_{q, u, n_a, s_a} \oslash \overleftarrow{\bm{\tau}}^{x_{\text F}}_{q, u, n_a, s_a})$. Similar to (\ref{LMMSEC}), with the message $\Delta_{f^{\mathbf x_{\text F}}_{u}
}^{\mathbf x_{u}}=\mathcal{CN}(\mathbf x_{u} \mid \overrightarrow{\mathbf{x}}_{u},\operatorname{diag}(\overrightarrow{\bm{\tau}}_{u}^x))$ as prior and (\ref{effecxF}) as likelihood, the LMMSE estimate of $\mathbf x_{u}$ is given by
\begin{align}
\label{postx}
	\hat{\mathbf x}_{u}=\bm{\Sigma}_{\mathbf x_{u}}(\mathbf{F}_{\text b}^{\mathrm{H}}\operatorname{diag}^{-1}(\overleftarrow{\bar{\bm{\tau}}}^{x_{\text F}}_{u}) \overleftarrow{\bar{\mathbf{x}}}^{\text F}_{u} + \operatorname{diag}^{-1}(\overrightarrow{\bm{\tau}}_{u}^x) \overrightarrow{\mathbf{x}}_{u}),
\end{align}
where 
\begin{align}
	\bm{\Sigma}_{\mathbf x_{u}}=(\mathbf{F}_{\text b}^{\mathrm{H}}\operatorname{diag}^{-1}(\overleftarrow{\bar{\bm{\tau}}}^{x_{\text F}}_{u})\mathbf{F}_{\text b} + \operatorname{diag}^{-1}(\overrightarrow{\bm{\tau}}_{u}^x))^{-1}.\label{posttaux}
\end{align}
The posterior distribution of $\mathbf x_{u}$ can also be written as $\Delta_{\mathbf x_{u}}=\mathcal{CN}(\mathbf x_{u} \mid \hat{\mathbf x}_{u}, \operatorname{diag}(\bm{\tau}_{u}^{x}))$, where $\bm{\tau}_{u}^{x}=\operatorname{diag}(\bm{\Sigma}_{\mathbf x_{u}})$. Then, we can get 
\begin{align}
\label{xtoxF}
	\Delta^{f^{\mathbf x_{\text F}}_{u}}_{\mathbf x_{u}}
	\propto
	\frac{\Delta_{\mathbf x_{u}}}{\Delta_{f^{\mathbf x_{\text F}}_{u}
		}^{\mathbf x_{u}}}
	\propto
	\mathcal{CN}(\mathbf x_{u} \mid \overleftarrow{\mathbf{x}}_{u},\operatorname{diag}(\overleftarrow{\bm{\tau}}_{u}^x)),
\end{align}
where $\overleftarrow{\bm{\tau}}_{u}^x=1\oslash(1\oslash\bm{\tau}_{u}^{x}-1\oslash\overrightarrow{\bm{\tau}}_{u}^x)$ and $\overleftarrow{\mathbf{x}}_{u}=\overleftarrow{\bm{\tau}}_{u}^x\odot(\hat{\mathbf x}_{u}\oslash\bm{\tau}_{u}^{x}-\overrightarrow{\mathbf{x}}_{u}\oslash\overrightarrow{\bm{\tau}}_{u}^x)$. With the update of $\Delta_{f^{\mathbf x_{\text F}}_{u}}^{\mathbf x_{u}}=\Delta^{f^{\mathbf x_{\text F}}_{u}}_{\mathbf x_{u}}$, we can compute $\Delta^{f^{\mathbf x_{\text F}}_{u}
}_{\mathbf x^{\text F}_{u}} = \mathcal{CN}(\mathbf x^{\text F}_{u}\mid \overrightarrow{\bar{\mathbf{x}}}^{\text F}_{u}, \operatorname{diag}(\overrightarrow{\bar{\bm{\tau}}}^{x_{\text F}}_{u}))$, where $\overrightarrow{\bar{\mathbf{x}}}^{\text F}_{u}=\mathbf{F}_{\text b} \overrightarrow{\mathbf{x}}_{u}$ and $\overrightarrow{\bar{\bm{\tau}}}^{x_{\text F}}_{u} = \operatorname{diag}(\mathbf{F}_{\text b} \operatorname{diag}(\overrightarrow{\bm{\tau}}_{u}^x)
\mathbf{F}_{\text b}^{\mathrm{H}})$. Now, the posterior distribution of $\mathbf x^{\text F}_{u}$ can be updated as
\begin{align}
\label{DeltaxF}
	\Delta_{\mathbf x_u}^{\text F} \propto
	\Delta^{f^{\mathbf x_{\text F}}_{u}
	}_{\mathbf x^{\text F}_{u}}\Delta_{f^{\mathbf x_{\text F}}_{u}
}^{\mathbf x^{\text F}_{u}} \propto
	\mathcal{CN}(\mathbf x^{\text F}_{u}\mid \hat{\mathbf{x}}^{\text F}_{u}, \operatorname{diag}(\bm{\tau}^{x_{\text F}}_{u})),
\end{align}
where $\bm{\tau}^{x_{\text F}}_{u}=1\oslash(1\oslash\overrightarrow{\bar{\bm{\tau}}}^{x_{\text F}}_{u} + 1\oslash \overleftarrow{\bar{\bm{\tau}}}^{x_{\text F}}_{u})$ and $\hat{\mathbf{x}}^{\text F}_{u}=\bm{\tau}^{x_{\text F}}_{u}\odot(\overrightarrow{\bar{\mathbf{x}}}^{\text F}_{u}\oslash \overrightarrow{\bar{\bm{\tau}}}^{x_{\text F}}_{u} + \overleftarrow{\bar{\mathbf{x}}}^{\text F}_{u}\oslash \overleftarrow{\bar{\bm{\tau}}}^{x_{\text F}}_{u})$. Finally, the message $\Delta_{f^{\text{P}}_{q, u, n_a, s_a}}^{\mathbf x^{\text F}_{u}}$ can be updated by
\begin{align}
\label{DeltaxF1}
	\Delta_{f^{\text{P}}_{q, u, n_a, s_a}}^{\mathbf x^{\text F}_{u}}
	&\propto 
	\frac{\Delta_{\mathbf x_u}^{\text F}}{\Delta^{f^{\text{P}}_{q, u, n_a, s_a}}_{\mathbf x^{\text F}_{u}}} \nonumber \\
	&\propto 
	 \mathcal{CN}(\mathbf x^{\text F}_{u}\mid \overrightarrow{\mathbf{x}}^{\text F}_{q, u, n_a, s_a}, \operatorname{diag}(\overrightarrow{\bm{\tau}}^{x_{\text F}}_{q, u, n_a, s_a})),
\end{align}
where $\overrightarrow{\bm{\tau}}^{x_{\text F}}_{q, u, n_a, s_a}=1\oslash(1\oslash\bm{\tau}^{x_{\text F}}_{u}-1\oslash\overleftarrow{\bm{\tau}}^{x_{\text F}}_{q, u, n_a, s_a})$ and $\overrightarrow{\mathbf{x}}^{\text F}_{q, u, n_a, s_a}=\overrightarrow{\bm{\tau}}^{x_{\text F}}_{q, u, n_a, s_a}\odot(\hat{\mathbf{x}}^{\text F}_{u}\oslash\bm{\tau}^{x_{\text F}}_{u}-\overleftarrow{\mathbf{x}}^{\text F}_{q, u, n_a, s_a}\oslash\overleftarrow{\bm{\tau}}^{x_{\text F}}_{q, u, n_a, s_a})$. 

Based on the above described procedures, we outline the main steps of the proposed centralized AEP in Algorithm 2. Note that the computation of $\hat{\mathbf x}_{u}$ and $\bm{\tau}_{u}^{x}$  involves costly matrix inversion, which may impose significantly computational burden on the central node. To mitigate this challenge, we employ the
covariance-free method (CFM) \cite{c11}. It transforms the matrix inversions into solving the linear equations, which can be computed in parallel using efficient algorithms only relying on matrix multiplication, such as the conjugate gradient method utilized in CFM. In addition, we recommend damping the messages that follow the interference cancellation step to prevent the algorithm from diverging. Let message $\Delta^{f^{\mathbf g^{\prime}}_{n_a,s_a}}_{\mathbf d^{\text{B}}_{q, n_a, s_a}}$ be computed by (\ref{ICC1}), and then the new message can be updated as
\begin{align}
	(\Delta^{f^{\mathbf g}_{n_a,s_a}}_{\mathbf d^{\text{B}}_{q, n_a, s_a}})_{\text{new}}
	\propto
	(\Delta^{f^{\mathbf g^{\prime}}_{n_a,s_a}}_{\mathbf d^{\text{B}}_{q, n_a, s_a}})^{\eta_{\epsilon}}
	((\Delta^{f^{\mathbf g}_{n_a,s_a}}_{\mathbf d^{\text{B}}_{q, n_a, s_a}})_{\text{old}})^{1-\eta_{\epsilon}},
\end{align}
where $\eta_{\epsilon}$ is the damping factor and we set $\eta_{\epsilon}=0.5$ in the simulations. The message $\Delta_{\mathbf d^{\text{P}}_{q, u, n_a, s_a}}^{f^{\text W}_{q,n_a,s_a}}$ can be damped similarly. Besides, the MRF-BGM-AMP and GAMP detector can also be adopted in the end of the centralized AEP for further improving performance.

Next, we describe the complexity of the proposed algorithm which consists of initial phase and refinement phase. In the initial phase, the MRF-BGM-AMP and GAMP detector is adopted, resulting in the complexity of $\mathcal{O}(S_aUM_{\mathrm{p}}N_{\mathrm{p}}(Q_{\mathrm I}+1)(L+1)N_zN_y + S_aN_zN_y\left|\mathcal{U}_a\right|(MN)^2)$. Note that due to the BEM modeling, the matrix-vector products involved in the refinement phase can be implemented by fast Fourier transform. Consequently, the complexity is primarily governed by the calculations of $\hat{\mathbf{c}}_{q,u,s_a}^{n_a}$, $\tau_{q,u,s_a}^{c_{n_a}}$, $\hat{\mathbf x}_{u}$, and $\bm{\tau}_{u}^{x}$. Computations for $\hat{\mathbf{c}}_{q,u,s_a}^{n_a}$ and $\tau_{q,u,s_a}^{c_{n_a}}$ involve matrix inversion, resulting in the complexity of $\mathcal{O}(S_aN_zN_y\left|\mathcal{U}_a\right|(Q+1)(L+1)^3)$. Typically, the number of channel taps $L+1$ is sufficiently small, e.g., $L=2$ in NTN-TDL-D \cite{3gpp}, and hence we can perform the matrix inversion directly. For computing $\hat{\mathbf x}_{u}$, and $\bm{\tau}_{u}^{x}$, we adopt the CFM which only needs the matrix multiplication, and then can also be implemented by FFT. As a result, it induces the complexity of $\mathcal{O}(\left|\mathcal{U}_a\right|(K_p+1)MN\log(MN))$, where $K_p$ is the number of probe vectors used in CFM.

\begin{algorithm}[hbt!]
	\footnotesize
	\caption{Centralized AEP}\label{Centralized AEP}
	\begin{algorithmic}[1]
		\REQUIRE{Recieved symbols $\mathbf{y}_{n_a,s_a}^{\prime}$; the maximum number of iterations $T_{\text {out}}$, $T_{\text{c}}$, $T_{\text{x}}$; the termination threshold $\eta_{\text c}$, $\eta_{\text x}$}.
		\ENSURE The refined time-varying channel $\hat h_{u,s_a}^{n_a}[n, l^{\prime}]$ and data symbols $\hat{\mathbf{x}}_u$.
		\STATE Get initial CE $\hat{\mathbf{c}}_{q,u,s_a}^{\text{r}_{n_a}}$,$\tau_{q,u,s_a}^{c\text r_{n_a}}$, and $\sigma^2_{s_a}$ by Algorithm 1
		\STATE Perform device identification via (\ref{DI})
		\STATE Get initial data symbols and variance $\hat{\mathbf{x}}_u^{\mathrm I}$ and $\bm{\tau}_u^{x_{\mathrm I}}$ by solving (\ref{intialSD})
		\STATE \textbf{Initialization}:
		$\forall q,u,n_a,s_a$: $\overleftarrow{\mathbf{c}}_{q,u,s_a}^{n_a}= \hat{\mathbf{c}}_{q,u,s_a}^{\text{r}_{n_a}}$,
		$\overleftarrow{\tau}_{q,u,s_a}^{c_{n_a}}=\tau_{q,u,s_a}^{c\text r_{n_a}}$,
		$\overleftarrow{\mathbf{c}}^{\text F}_{q,u,n_a, s_a}=\mathbf{F}_{1:L+1} \overleftarrow{\mathbf{c}}_{q,u,s_a}^{n_a}$, 
		$\overleftarrow{{\tau}}^{c_{\text F}}_{q,u,n_a, s_a}=\overleftarrow{\tau}_{q,u,s_a}^{c_{n_a}}$,
		$\overrightarrow{\mathbf{x}}_{u}=\hat{\mathbf{x}}_u^{\mathrm I}$, $\overrightarrow{\bm{\tau}}_{u}^x=\bm{\tau}_u^{x_{\mathrm I}}$,
		$\overleftarrow{\bar{\mathbf{x}}}^{\text F}_{u}=\mathbf 0$, $\overleftarrow{\bar{\bm{\tau}}}^{x_{\text F}}_{u}=\bm{\infty}$, $\overrightarrow{\bar{\mathbf{x}}}^{\text F}_{u}=\mathbf{F}_{\text b} \overrightarrow{\mathbf{x}}_{u}$, $\overrightarrow{\bar{\bm{\tau}}}^{x_{\text F}}_{u} = \operatorname{diag}(\mathbf{F}_{\text b} \operatorname{diag}(\overrightarrow{\bm{\tau}}_{u}^x)
		\mathbf{F}_{\text b}^{\mathrm{H}})$
		\FOR{$t_{\text{out}}=1$ to $T_{\text{out}}$}
		\STATE 	$\forall q,u,n_a,s_a$: Compute $\Delta_{\mathbf x_u}^{\text F}$ and $\Delta_{f^{\text{P}}_{q, u, n_a, s_a}}^{\mathbf x^{\text F}_{u}}$ via (\ref{DeltaxF}) and (\ref{DeltaxF1})
		\STATE $\backslash \backslash$ \textbf{Refine channel}
		\FOR{$t_{\text{c}}=1$ to $T_{\text{c}}$}
		\STATE $\forall q,u,n_a,s_a$: Compute $\Delta^{f^{\text{P}}_{q, u, n_a, s_a}}_{\mathbf c_{q,u,n_a, s_a}^{\text F}}$ via (\ref{ICstart})-(\ref{ICend}) and (\ref{ICc})
		\STATE $\forall q,u,n_a,s_a$: Compute $\hat{\mathbf{c}}_{q,u,s_a}^{n_a}(t_{\text{c}}+1)$ and $\tau_{q,u,s_a}^{c_{n_a}}$ via (\ref{LMMSEC}) and (\ref{overleftarrow_tau})
		\STATE $\forall q,u,n_a,s_a$: Update $\Delta^{\mathbf{c}_{q,u,s_a}^{n_a}}_{f^{\mathbf c_{\text F}}_{q, u, n_a, s_a}}$ via (\ref{ctocF})
		\STATE $\forall q,u,n_a,s_a$: Update $\overleftarrow{\mathbf{c}}^{\text F}_{q,u,n_a, s_a}=\mathbf{F}_{1:L+1} \overleftarrow{\mathbf{c}}_{q,u,s_a}^{n_a}$ and $\overleftarrow{{\tau}}^{c_{\text F}}_{q,u,n_a, s_a}=\overleftarrow{\tau}_{q,u,s_a}^{c_{n_a}}$
		\STATE \textbf{if} $\frac{\sum_{q,u,n_a,s_a}\left\|\hat{\mathbf{c}}_{q,u,s_a}^{n_a}(t_{\text{c}}+1)-\hat{\mathbf{c}}_{q,u,s_a}^{n_a}(t_{\text{c}})\right\|^2}{\sum_{q,u,n_a,s_a}\left\|\hat{\mathbf{c}}_{q,u,s_a}^{n_a}(t_{\text{c}})\right\|^2} \leq \eta_{\text c} $: break \textbf{end if}
		\ENDFOR
		\STATE $\forall q,u,n_a,s_a$: Compute $\Delta_{\mathbf c_{q,u,n_a, s_a}}^{{\text F}}$ via (\ref{DeltacF})
		\STATE $\backslash \backslash$ \textbf{Refine data symbols}
		\FOR{$t_{\text{x}}=1$ to $T_{\text{x}}$}
		\STATE $\forall q,u,n_a,s_a$: Compute $\Delta^{f^{\text{P}}_{q, u, n_a, s_a}}_{\mathbf x^{\text F}_{u}}$ via (\ref{ICstart})-(\ref{ICend}) and (\ref{ICx})
		\STATE $\forall u$: Compute $\Delta_{f^{\mathbf x_{\text F}}_{u}
		}^{\mathbf x^{\text F}_{u}}$ via (\ref{effecxF})
		\STATE $\forall u$: Compute $\hat{\mathbf x}_{u}(t_{\text{x}}+1)$ and $\bm{\tau}_{u}^{x}$ via (\ref{postx}) and (\ref{posttaux})
		\STATE $\forall u$: Update $\Delta_{f^{\mathbf x_{\text F}}_{u}
		}^{\mathbf x_{u}}$ via (\ref{xtoxF})
		\STATE $\forall u$: Update $\overrightarrow{\bar{\mathbf{x}}}^{\text F}_{u}=\mathbf{F}_{\text b} \overrightarrow{\mathbf{x}}_{u}$ and $\overrightarrow{\bar{\bm{\tau}}}^{x_{\text F}}_{u} = \operatorname{diag}(\mathbf{F}_{\text b} \operatorname{diag}(\overrightarrow{\bm{\tau}}_{u}^x)
		\mathbf{F}_{\text b}^{\mathrm{H}})$
		\STATE \textbf{if} $\frac{\sum_{u}\left\|\hat{\mathbf x}_{u}(t_{\text{x}}+1)-\hat{\mathbf x}_{u}(t_{\text{x}})\right\|^2}{\sum_{u} \left\|\hat{\mathbf x}_{u}(t_{\text{x}})\right\|^2} \leq \eta_{\text x}$: break \textbf{end if}
		\ENDFOR
		\ENDFOR 
		\STATE $\forall n,l^{\prime},u,n_a,s_a$: Reconstruct $\hat h_{u,s_a}^{n_a}[n, l^{\prime}]$ via (\ref{has})
		\STATE $\forall u$: Perform hard decision on $\hat{\mathbf x}_{u}$
	\end{algorithmic}
\end{algorithm}
\vspace{-0.5cm}
\subsection{Distributed Cooperative Mode}
Since the centralized mode imposes the huge computational burden on the central server, we next investigate the distributed method which allows for offloading the computations to edge satellites.
\subsubsection{Device identification}
In the distributed mode, the Algorithm 1 is firstly performed at each satellite independently to get the initial channel estimation $\hat c^{\mathrm I^{u,s_a}}_{i,j}$. To identify active devices, each satellite computes the channel energy $\sum_{i,j}\left|\hat c^{\mathrm I^{u,s_a}}_{i,j}\right|^2$, and broadcast it within the constellations. Then, the activity indicator of $u$-th device can be given by (\ref{DI}).
\subsubsection{Initial symbol detection}
Here, each satellite detect data symbols independently in the initial phase, and then we have the following relationship.
\begin{align}
\label{intialDSD}
\mathbf{y}^{\text{DD}}_{s_a}=\bar{\mathbf{H}}_{s_a} \mathbf{x} +  \mathbf{e}_{s_a},
\end{align}
where $\bar{\mathbf{H}}_{s_a}=[\bar{\mathbf{H}}_{0,s_a}^{\mathrm{T}},\dots,\bar{\mathbf{H}}_{N_zN_y-1,s_a}^{\mathrm{T}}]^{\mathrm{T}}$. The difference from (\ref{intialSD}) is that the channel matrix $\bar{\mathbf{H}}_{s_a}$ only involves the channels related to the $s_a$-th satellite. Then, the GAMP with uniform prior can be employed to get the initial guess of data symbols $\hat{\mathbf{x}}_{u,s_a}^{\mathrm I}$ for active devices and the corresponding variance $\bm{\tau}_{u,s_a}^{x_{\mathrm I}}$. 
\subsubsection{Jointly refine channel estimation and symbol detection}
To exploit diversity provided by the satellite constellations, we propose to share the soft information about data symbols among the $S_a$ satellites during the iteration process of the refinement algorithm. Hence, except for the previous defined auxiliary variables, we introduce $\mathbf x_{u,s_a} = \mathbf x_u$ and the corresponding factors $f_{u,s_a,s_a^{\prime}}^{\mathbf x}=\delta(\mathbf x_{u,s_a}-\mathbf x_{u,s_a^{\prime}})$, $s_a^{\prime}=\{0,\dots,S_a-1\}\backslash s_a$. 
Note that the previous derivations of IC module and CER module for centralized mode can also be applied here. However, there are two significant differences in the SDR module: the one is that the likelihoods from $S_a$ satellites can not be combined directly via (\ref{effecxF}), since the refinement algorithm is implemented independently at each satellite; the other is that the message from $\mathbf{x}_{u,s_a}$ to $f^{\mathbf x_{\text F}}_{u,s_a}$ should combine the soft information shared by other satellites. 

Firstly, by considering the likelihoods at the $s_a$-th satellite, we revise the message in (\ref{effecxF}) as
\begin{align}
\label{effecxFsa}
\Delta_{f^{\mathbf x_{\text F}}_{u,s_a}
}^{\mathbf x^{\text F}_{u,s_a}} 
&\propto
\prod_{q,n_a}\Delta^{f^{\text{P}}_{q, u, n_a,s_a}}_{\mathbf x^{\text F}_{u,s_a}} \nonumber \\
&\propto \mathcal{CN}(\mathbf x^{\text F}_{u,s_a}\mid \overleftarrow{\bar{\mathbf{x}}}^{\text F}_{u,s_a}, \operatorname{diag}(\overleftarrow{\bar{\bm{\tau}}}^{x_{\text F}}_{u,s_a})),
\end{align}
where $\overleftarrow{\bar{\bm{\tau}}}^{x_{\text F}}_{u,s_a}=1\oslash \sum_{q,n_a} (1\oslash \overleftarrow{\bm{\tau}}^{x_{\text F}}_{q, u, n_a,s_a})$ and $\overleftarrow{\bar{\mathbf{x}}}^{\text F}_{u,s_a}=\overleftarrow{\bar{\bm{\tau}}}^{x_{\text F}}_{u,s_a}\odot\sum_{q,n_a} (\overleftarrow{\mathbf{x}}^{\text F}_{q, u, n_a, s_a} \oslash \overleftarrow{\bm{\tau}}^{x_{\text F}}_{q, u, n_a, s_a})$. Besides, with the message shared by other satellites $\Delta^{f_{u,s_a,s_a^{\prime}}^{\mathbf x}}_{\mathbf x_{u,s_a}} = \mathcal{CN}(\mathbf x_{u,s_a}\mid \overrightarrow{\bar{\mathbf{x}}}_{u,s_a,s_a^{\prime}}, \operatorname{diag}(\overrightarrow{\bar{\bm{\tau}}}^{x}_{u,s_a,s_a^{\prime}}))$, the message from $\mathbf{x}_{u,s_a}$ to $f^{\mathbf x_{\text F}}_{u,s_a}$ can be updated as 
\begin{align}
\label{xtoxFsa}
	\Delta_{f^{\mathbf x_{\text F}}_{u,s_a}
	}^{\mathbf x_{u,s_a}}
	&\propto
	\Delta^{f^{\mathbf x_{\text F}}_{u,s_a}}_{\mathbf x_{u,s_a}}
	\prod_{s_a^{\prime}} \Delta^{f_{u,s_a,s_a^{\prime}}^{\mathbf x}}_{\mathbf x_{u,s_a}} \nonumber \\
	&\propto
	\mathcal{CN}(\mathbf x_{u,s_a} \mid \overrightarrow{\mathbf{x}}_{u,s_a},\operatorname{diag}(\overrightarrow{\bm{\tau}}_{u,s_a}^x)),
\end{align}
where $\overrightarrow{\bm{\tau}}_{u,s_a}^x=1\oslash(1\oslash\overleftarrow{\bm{\tau}}_{u,s_a}^x + \sum_{s_a^{\prime}} (1\oslash \overrightarrow{\bar{\bm{\tau}}}^{x}_{u,s_a,s_a^{\prime}}))$ and $\overrightarrow{\mathbf{x}}_{u,s_a}=\overrightarrow{\bm{\tau}}_{u,s_a}^x \odot (\overleftarrow{\mathbf{x}}_{u,s_a}\oslash \overleftarrow{\bm{\tau}}_{u,s_a}^x + \sum_{s_a^{\prime}}( \overrightarrow{\bar{\mathbf{x}}}_{u,s_a,s_a^{\prime}} \oslash \overrightarrow{\bar{\bm{\tau}}}^{x}_{u,s_a,s_a^{\prime}}))$. Then, the message transmitted to other satellites can be obtained as
\begin{align}
\label{transmitSa}
	\Delta_{f_{u,s_a,s_a^{\prime}}^{\mathbf x}}^{\mathbf x_{u,s_a}}
	&\propto
	\frac{\Delta_{f^{\mathbf x_{\text F}}_{u,s_a}
		}^{\mathbf x_{u,s_a}}}{\Delta^{f_{u,s_a,s_a^{\prime}}^{\mathbf x}}_{\mathbf x_{u,s_a}}} \nonumber \\
	&\propto
	\mathcal{CN}(\mathbf x_{u,s_a}\mid \overleftarrow{\bar{\mathbf{x}}}_{u,s_a,s_a^{\prime}}, \operatorname{diag}(\overleftarrow{\bar{\bm{\tau}}}^{x}_{u,s_a,s_a^{\prime}})),
\end{align}
where $\overleftarrow{\bar{\bm{\tau}}}^{x}_{u,s_a,s_a^{\prime}}=1\oslash(1\oslash\overrightarrow{\bm{\tau}}_{u,s_a}^x - 1\oslash\overrightarrow{\bar{\bm{\tau}}}^{x}_{u,s_a,s_a^{\prime}})$ and $\overleftarrow{\bar{\mathbf{x}}}_{u,s_a,s_a^{\prime}}=\overleftarrow{\bar{\bm{\tau}}}^{x}_{u,s_a,s_a^{\prime}}\odot(\overrightarrow{\mathbf{x}}_{u,s_a}\oslash\overrightarrow{\bm{\tau}}_{u,s_a}^x-\overrightarrow{\bar{\mathbf{x}}}_{u,s_a,s_a^{\prime}} \oslash \overrightarrow{\bar{\bm{\tau}}}^{x}_{u,s_a,s_a^{\prime}})$.
Finally, we can update $\Delta^{f_{u,s_a,s_a^{\prime}}^{\mathbf x}}_{\mathbf x_{u,s_a}}$ as
\begin{align}
\label{ReceiveSa}
	\Delta^{f_{u,s_a,s_a^{\prime}}^{\mathbf x}}_{\mathbf x_{u,s_a}} 
	\propto
	\int_{\mathbf x_{u,s_a^{\prime}}} f_{u,s_a,s_a^{\prime}}^{\mathbf x} \Delta_{f_{u,s_a,s_a^{\prime}}^{\mathbf x}}^{\mathbf x_{u,s_a^{\prime}}},
\end{align}
where $\overrightarrow{\bar{\mathbf{x}}}_{u,s_a,s_a^{\prime}}=\overleftarrow{\bar{\mathbf{x}}}_{u,s_a^{\prime},s_a}$ and  $\overrightarrow{\bar{\bm{\tau}}}^{x}_{u,s_a,s_a^{\prime}}=\overleftarrow{\bar{\bm{\tau}}}^{x}_{u,s_a^{\prime},s_a}$. Other derivations of SDR module in centralized mode can be reused by adding the index $s_a$, which represents the computations are implemented at the $s_a$-th satellite. 

The proposed distributed AEP algorithm is summarized in Algorithm \ref{Distributed AEP}. In Line 5, $T_{\text{ex}}$ represents the number of iterations during which the soft information of data symbols is exchanged.
	Line 6 indicates that the subsequent computations are individually performed at each satellite, which is executed in parallel for practical deployment. Lines 21 and 24 handle the exchange of soft information about data symbols within the satellite constellations. Totally, the soft information about data symbols is exchanged $T_{\text{ex}}+1$ times in Algorithm \ref{Distributed AEP}.  Furthermore, the damping and CFM method can also be adopted here. The complexity analysis for the distributed AEP is similar to that of the centralized AEP, except that the dominant computations are offloaded to edge satellites. Therefore, for a single satellite, the complexities of the initial phase and the refinement phase are in the order of $\mathcal{O}(UM_{\mathrm{p}}N_{\mathrm{p}}(Q_{\mathrm I}+1)(L+1)N_zN_y + N_zN_y\left|\mathcal{U}_a\right|(MN)^2)$ and $\mathcal{O}(N_zN_y\left|\mathcal{U}_a\right|(Q+1)(L+1)^3 + \left|\mathcal{U}_a\right|(K_p+1)MN\log(MN))$, respectively.

\begin{algorithm}[hbt!]
	\footnotesize
	\caption{{Distributed AEP}}\label{Distributed AEP}
	\begin{algorithmic}[1]
		\REQUIRE{Recieved symbols $\mathbf{y}_{n_a,s_a}^{\prime}$; the maximum number of iterations $T_{\text{ex}}$, $T_{\text {out}}$, $T_{\text{c}}$, $T_{\text{x}}$; the termination threshold $\eta_{\text c}$, $\eta_{\text x}$}.
		\ENSURE The refined time-varying channel $\hat h_{u,s_a}^{n_a}[n, l^{\prime}]$ and data symbols $\hat{\mathbf{x}}_u$.
		\STATE Get initial CE $\hat{\mathbf{c}}_{q,u,s_a}^{\text{r}_{n_a}}$,$\tau_{q,u,s_a}^{c\text r_{n_a}}$, and $\sigma^2_{s_a}$ by Algorithm 1
		\STATE Perform device identification via (\ref{DI})
		\STATE Get initial data symbols and variance $\hat{\mathbf{x}}_{u,s_a}^{\mathrm I}$ and $\bm{\tau}_{u,s_a}^{x_{\mathrm I}}$ by solving (\ref{intialDSD})
		\STATE \textbf{Initialization}:
		Similar with line 4 in Algorithm 2 except for 
		$\forall u,s_a,s_a^{\prime}$:
		$\overrightarrow{\bar{\mathbf{x}}}_{u,s_a,s_a^{\prime}}=\mathbf 0$, $\overrightarrow{\bar{\bm{\tau}}}^{x}_{u,s_a,s_a^{\prime}}=\bm{\infty}$
		\FOR{$t_{\text{ex}}=1$ to $T_{\text{ex}}$}
		\FOR{$s_a=0$ to $S_a-1$}
		\FOR{$t_{\text{out}}=1$ to $T_{\text{out}}$}
		\STATE 	$\forall q,u,n_a$: Compute $\Delta_{\mathbf x_{u,s_a}}^{\text F}$ and $\Delta_{f^{\text{P}}_{q, u, n_a, s_a}}^{\mathbf x^{\text F}_{u,s_a}}$ via (\ref{DeltaxF}) and (\ref{DeltaxF1})
		\STATE Refine channel using lines 8-15 of Algorithm 2 with fixed $s_a$
		\STATE $\backslash \backslash$ \textbf{Refine data symbols}
		\FOR{$t_{\text{x}}=1$ to $T_{\text{x}}$}
		\STATE $\forall q,u,n_a$: Compute $\Delta^{f^{\text{P}}_{q, u, n_a, s_a}}_{\mathbf x^{\text F}_{u,s_a}}$ via (\ref{ICstart})-(\ref{ICend}) and (\ref{ICx})
		\STATE $\forall u$: Compute $\Delta_{f^{\mathbf x_{\text F}}_{u,s_a}
		}^{\mathbf x^{\text F}_{u,s_a}}$ via (\ref{effecxFsa})
		\STATE $\forall u$: Compute $\hat{\mathbf x}_{u,s_a}(t_{\text{x}}+1)$ and $\bm{\tau}_{u,s_a}^{x}$ via (\ref{postx}) and (\ref{posttaux})
		\STATE $\forall u$: Update $\Delta_{f^{\mathbf x_{\text F}}_{u,s_a}
		}^{\mathbf x_{u,s_a}}$ via (\ref{xtoxFsa})
		\STATE $\forall u$: Update $\overrightarrow{\bar{\mathbf{x}}}^{\text F}_{u,s_a}=\mathbf{F}_{\text b} \overrightarrow{\mathbf{x}}_{u,s_a}$ and $\overrightarrow{\bar{\bm{\tau}}}^{x_{\text F}}_{u,s_a} = \operatorname{diag}(\mathbf{F}_{\text b} \operatorname{diag}(\overrightarrow{\bm{\tau}}_{u,s_a}^x)
		\mathbf{F}_{\text b}^{\mathrm{H}})$
		\STATE \textbf{if} $\frac{\sum_{u}\left\|\hat{\mathbf x}_{u,s_a}(t_{\text{x}}+1)-\hat{\mathbf x}_{u,s_a}(t_{\text{x}})\right\|^2}{\sum_{u} \left\|\hat{\mathbf x}_{u,s_a}(t_{\text{x}})\right\|^2} \leq \eta_{\text x}$: break \textbf{end if}
		\ENDFOR
		\ENDFOR 
		\ENDFOR
		\STATE $\forall u,s_a,s_a^{\prime}$: Compute
		$\Delta_{f_{u,s_a,s_a^{\prime}}^{\mathbf x}}^{\mathbf x_{u,s_a}}$ and $\Delta^{f_{u,s_a,s_a^{\prime}}^{\mathbf x}}_{\mathbf x_{u,s_a}}$ via (\ref{transmitSa}) and (\ref{ReceiveSa})
		\ENDFOR
		\STATE $\forall t,l^{\prime},u,n_a,s_a$: Reconstruct $\hat h_{u,s_a}^{n_a}[n, l^{\prime}]$ via (\ref{has})
		\STATE $\forall u$: $\hat{\mathbf x}_{u}=1\oslash(\sum_{s_a}1\oslash\bm{\tau}_{u,s_a}^{x})\odot(\sum_{s_a}\hat{\mathbf x}_{u,s_a}\oslash \bm{\tau}_{u,s_a}^{x})$
		\STATE $\forall u$: Perform hard decision on $\hat{\mathbf x}_{u}$
	\end{algorithmic}
\end{algorithm}

\section{Numerical Results} 
\label{V}
\begin{table}
	\scriptsize
	\caption{Simulation Parameters}
	\vspace{-.4cm}
	\begin{center}
		\begin{tabular}{|c|c|}
			\hline Parameter & Values \\
			\hline Channel Model & NTN-CDL-A \\
			\hline Carrier frequency & 2 GHz \\
			\hline Modulation Scheme & QPSK \\
			\hline Orbit altitude & 600 km \\
			\hline Spot beam radius & 200 km \\
			\hline Satellite velocity & 7.5622 km/s \\
			\hline Device velocity & $0\sim500$ km/h\\
			\hline Subcarrier spacing $\Delta f$ & 240 kHz \\
			\hline Frame size $(M, N)$ & $(32, 15)$ \\
			\hline Number of antennas $N_z \times N_y$ & $8 \times 8$\\
			\hline
		\end{tabular}
		\label{spa}
	\end{center}
\vspace{-.3cm}
\end{table}
This section presents simulation results to validate the effectiveness of the proposed algorithms, with the NTN-CDL-A channel model \cite{3gpp} recommended by 3GPP for the experiments. We consider a typical random access scenario in the non-terrestrial networks \cite{c13}, where $U=50$ potential devices are randomly distributed within the coverage area of the satellites. This area is defined as a circle with the 200 km radius. The satellites themselves are positioned evenly along the circumference of a larger circle with the 400 km radius, at the altitude of 600 km. The main parameters are summarized in Table \ref{spa}. The activity error rate (AER), the normalized mean squared error (NMSE), and the symbol error rate (SER) are adopted as the performance metrics, defined as $\text{AER}=\frac{1}{U}\sum_u \mathbb{I}\{\hat{\lambda}_u\neq \lambda_u\}$, $\text{NMSE}= \frac{\sum_{n, l^{\prime},n_a,u,s_a} \left|h_{u,s_a}^{n_a}[n, l^{\prime}]-\hat h_{u,s_a}^{n_a}[n, l^{\prime}]\right|^2}{\sum_{n, l^{\prime},n_a,u,s_a} \left|h_{u,s_a}^{n_a}[n, l^{\prime}]\right|^2}$, and $\text{SER}= \frac{\sum_{u}\sum_{m\in \mathcal{S}_d} \mathbb{I}\{x_u[m]\neq\hat x_u[m]\} }{U\left|\mathcal{S}_d\right|}$,
where $\mathcal{S}_d$ contains the indexes of the data symbols, assuming the inactive devices transmit zeroes. Then, a symbol is detected correctly only when both the device activity and the symbol itself are judged correctly. In addition, the signal-to-noise ratio is defined as $\text{SNR}=10\log_{10}\frac{\left\|\mathbf Z_{s_a}\right\|_{\mathrm F}^2}{M_{\mathrm p}N_{\mathrm p}N_zN_y\sigma^2_{s_a}}$. 
Two benchmarks are adopted for performance comparison. The first, ConvSBL-GAMP + MMSE, employs the ConvSBL-GAMP \cite{c11} for joint device identification and channel estimation, followed by symbol detection using the MMSE equalizer \cite{c22}. The second, Oracle-LS, leverages perfect CSI to detect data symbols based on the least squares principle \cite{c23}.
These benchmarks are evaluated for two modes: (i) centralized cooperative mode where the central server processes all received signals, leveraging the collective information for enhanced performance. (ii) non-cooperative mode where satellites perform the task independently and without signal sharing.

\begin{figure*}
	\centering
	\captionsetup{font={small}}
	\setlength{\belowcaptionskip}{-.6cm}
	\subfigure[Performance of channel estimation.]{
		\label{SNR_NMSE}
		\begin{minipage}{5.7cm}
			\includegraphics[width=\textwidth]{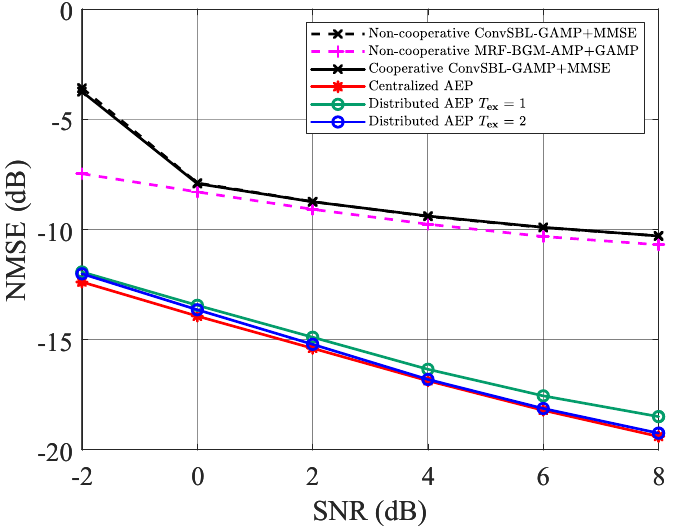} \\
			\vspace{-0.2cm}	
		\end{minipage}
	}
	\subfigure[Performance of symbol detection.]{
		\label{SNR_SER}
		\begin{minipage}{5.725cm}
			\includegraphics[width=\textwidth]{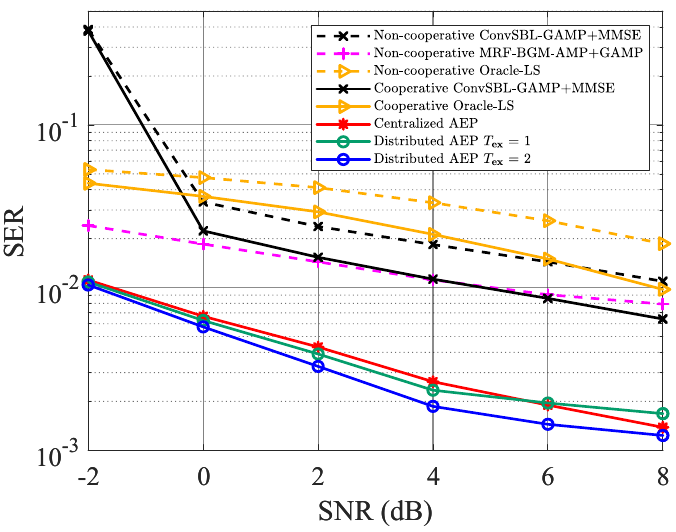} \\
			\vspace{-0.2cm}	
		\end{minipage}
	}
	\subfigure[Performance of device identification.]{
		\label{SNR_AER}
		\begin{minipage}{5.7cm}
			\includegraphics[width=\textwidth]{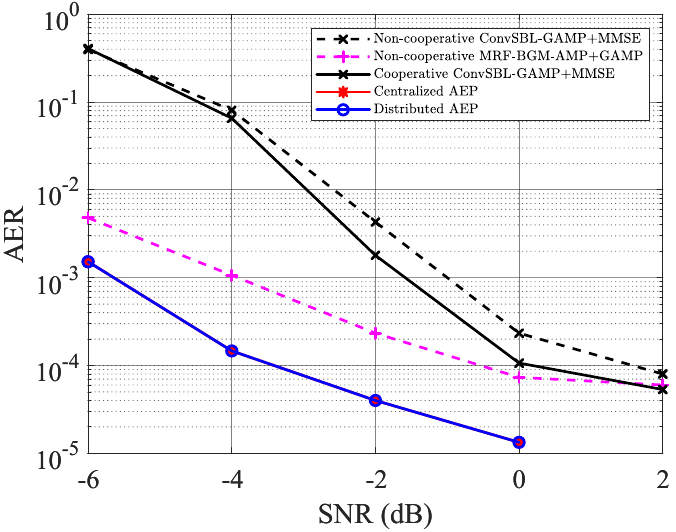} \\
			\vspace{-0.2cm}	
		\end{minipage}
	}
	\vspace{-0.2cm}
	\caption{Performance comparison among different schemes given different SNR values, where $S_a=2$, $p_{\lambda}=0.1$, and $\rho=0.4$.} 
	\label{SNRPer}
\end{figure*} 
Fig. \ref{SNRPer} depicts the NMSE, SER, and AER as the functions of SNR achieved by various algorithms, given the number of satellites $S_a=2$, device activity $p_{\lambda}=0.1$, and pilot overhead $\rho=0.4$. The BEM orders are set to $Q_I=4$ and $Q_R=8$ unless otherwise specified. Additionally, we include results for non-cooperative MRF-BGM-AMP+GAMP, which is the initial phase of the distributed AEP. It is observed that the cooperative schemes achieve superior performance compared to their non-cooperative counterparts, with the proposed cooperative algorithms significantly outperforming the benchmarks. For instance, in Fig. \ref{SNR_NMSE}, when the SNR is 4 dB, the proposed centralized AEP outperforms ConvSBL-GAMP by around 6 dB in terms of NMSE. Similarly, Fig. \ref{SNR_SER} indicates that at the SER of 0.01, both centralized and distributed AEP outperform the cooperative ConvSBL-GAMP+MMSE and Oracle-LS by approximately 6 dB and 9 dB in terms of SNR, respectively. This is due to the benefit of joint channel estimation and symbol detection design. Fig. \ref{SNR_AER} shows that at the AER of 0.001, both the centralized and distributed AEP outperform the cooperative ConvSBL-GAMP+MMSE by about 4 dB.



Fig. \ref{SNR_NMSE} and Fig. \ref{SNR_SER} also illustrate that increasing the iterations of soft information exchange of data symbols in distributed AEP leads to reduced NMSE and SER, and only two exchanges ($T_{\text{ex}}=1$) are needed for achieving performance comparable to the centralized AEP; it is indicated that distributed AEP can efficiently offload the computational burden while requiring small signaling overhead. Moreover, the distributed AEP achieves a marginal gain over the centralized AEP when $T_{\text{ex}}=2$, which results from the different ways they exploit the spatial diversity. In the distributed AEP, the soft information related to data symbols are produced by the LMMSE denoiser and then aggregated from different satellites, which is possible to obtain a more precise likelihoods for data symbols. We adopt $T_{\text{ex}}=2$ for subsequent distributed AEP. 


\begin{figure}
	\centering
	\captionsetup{font={small}}
	\setlength{\belowcaptionskip}{-.6cm}
	\subfigure[Performance of channel estimation.]{
		\label{Q_NMSE}
		\begin{minipage}{7cm}
			\includegraphics[width=\textwidth]{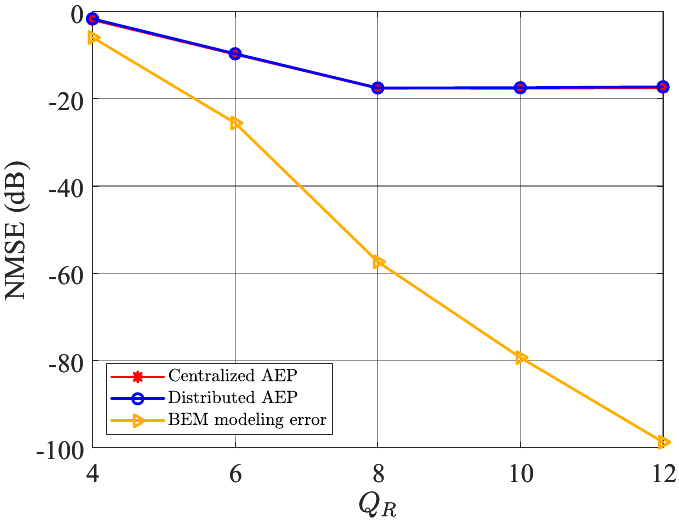} \\
		\vspace{-0.2cm}	
		\end{minipage}
	}
	\subfigure[Performance of symbol detection.]{
		\label{Q_SER}
		\begin{minipage}{7cm}
			\includegraphics[width=\textwidth]{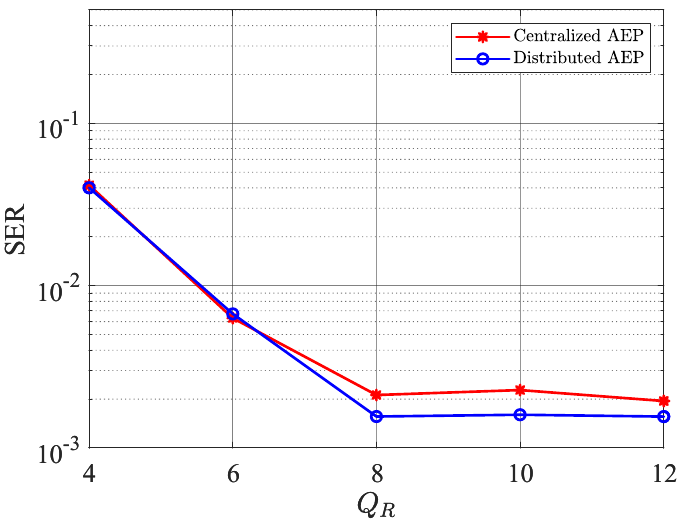} \\
		\vspace{-0.2cm}	
		\end{minipage}
	}
	\vspace{-0.2cm}
	\caption{The impact of BEM modeling accuracy, where SNR = 5 dB, $S_a=2$, $p_{\lambda}=0.1$, and $\rho=0.4$.} 
	\vspace{0.5cm}
	\label{QPer}
\end{figure} 
Fig. \ref{QPer} illustrates how the accuracy of BEM modeling influences the performance of the proposed algorithms by showing NMSE and SER as functions of the BEM order $Q_R$. The observations reveal that performance enhancements are evident when $Q_R\leq 8$, as higher $Q_R$ values lead to reduced BEM modeling errors. However, for $Q_R > 8$, performance gains plateau due to inter-user and inter-component interference, which limits further NMSE and SER improvements despite ongoing reductions in BEM modeling error. Notably, BEM with $Q_R =2\left\lceil 2 N \bar{\nu}_{\max}\right\rceil= 8$ suffices for precisely modeling TSL, achieving an NMSE of approximately -60 dB.

\begin{figure}
	\centering
	\captionsetup{font={small}}
	\setlength{\belowcaptionskip}{-.6cm}
	\subfigure[Performance of channel estimation.]{
		\label{Pilot_NMSE}
		\begin{minipage}{7cm}
			\includegraphics[width=\textwidth]{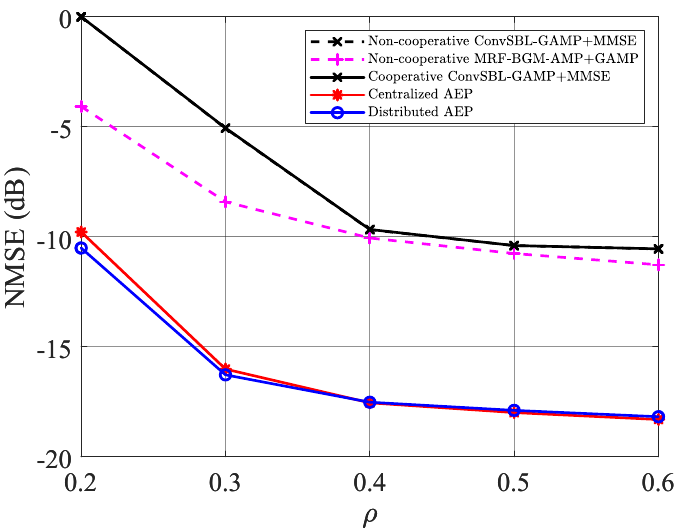} \\
		\vspace{-0.2cm}	
		\end{minipage}
	}
	\subfigure[Performance of symbol detection.]{
		\label{Pilot_SER}
		\begin{minipage}{7cm}
			\includegraphics[width=\textwidth]{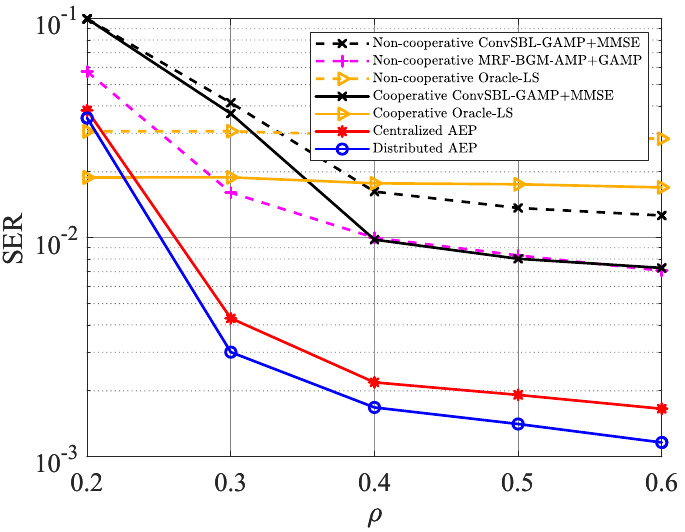} \\
		\vspace{-0.2cm}	
		\end{minipage}
	}
	\vspace{-0.2cm}
	\caption{Performance comparison among different schemes given different pilot overhead, where $S_a=2$, $p_{\lambda}=0.1$, and $\text{SNR}=5$ dB.} 
	\vspace{0.5cm}
	\label{PilotPer}
\end{figure}
\begin{figure*}
	\centering
	\captionsetup{font={small}}
	\setlength{\belowcaptionskip}{-.6cm}
	\subfigure[Symbol detection performance for different device acitivities, where $\text{SNR}=5$ dB.]{
		\label{Activity_SER}
		\begin{minipage}{5.7cm}
			\includegraphics[width=\textwidth]{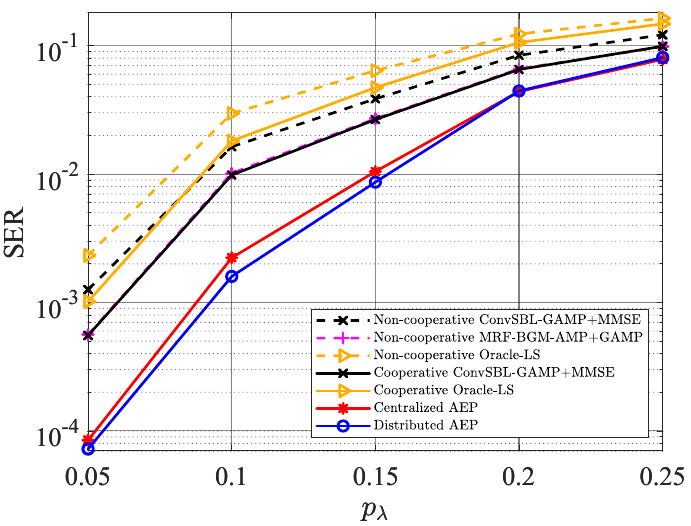} \\	
			\vspace{-0.2cm}
		\end{minipage}
	}
	\subfigure[Symbol detection performance for different number of satellites, where $\text{SNR}=5$ dB.]{
		\label{S_SER}
		\begin{minipage}{5.725cm}
			\includegraphics[width=\textwidth]{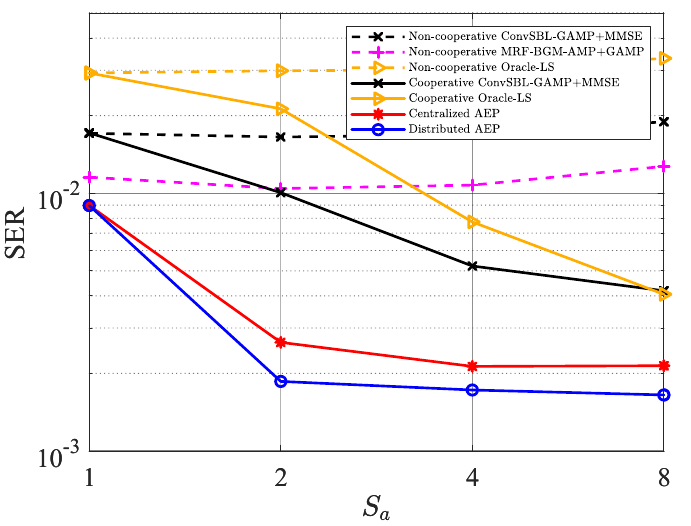} \\	
			\vspace{-0.2cm}
		\end{minipage}
	}
	\subfigure[Device identification performance for different number of satellites, where $\text{SNR}=-2$ dB.]{
		\label{S_AER}
		\begin{minipage}{5.725cm}
			\includegraphics[width=\textwidth]{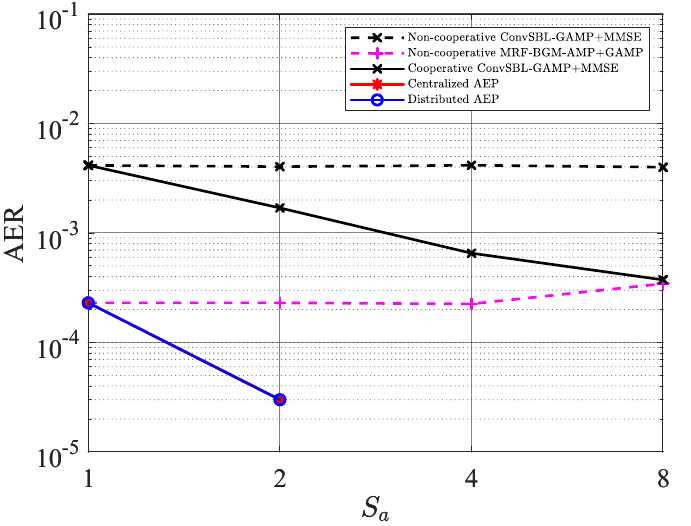} \\	
			\vspace{-0.2cm}
		\end{minipage}
	}
	\vspace{-0.2cm}
	\caption{Performance comparison for the system scalability among different schemes.} 
	\label{SPer}
\end{figure*}

In Fig. \ref{PilotPer}, we evaluate the performance of different schemes with varying pilot overhead, given the number of satellites $S_a=2$, device activity $p_{\lambda}=0.1$, and $\text{SNR}=5$ dB. 
Fig. \ref{Pilot_SER} reveals that at a pilot overhead of 0.2, Oracle-LS outperforms other methods in SER, benefiting from perfect channel knowledge for symbol detection. In contrast, other methods struggle with accurate channel estimation at such low pilot overhead, as shown in Fig. \ref{Pilot_NMSE}. However, when pilot overhead exceeds 0.3, our proposed cooperative algorithms demonstrate remarkable improvements over baselines. This enhancement results from more accurate initial channel estimation, facilitating the initial symbol detection and refinement phase of our algorithms. For instance, to achieve an SER of 0.01, both the centralized and distributed AEP require 0.15 less pilot overhead than the cooperative ConvSBL-GAMP+MMSE, indicating their higher spectral efficiency.

In Fig. \ref{SPer}, we investigate the scalability of the system. Fig. \ref{Activity_SER} plots the SER as a function of the device activity, given the number of satellites $S_a=2$, pilot overhead $\rho=0.4$, and $\text{SNR}=5$ dB. It is observed that the proposed centralized AEP and distributed AEP achieves the best performance. Note that the SER of all the schemes deteriorates obviously when
$p_{\lambda}$ is large. This is because the limited size of the receiving antenna array leads to decreased spatial separation as the number of active devices increases. Then, the inter-user interference and inter-component interference of BEM also become severer in the spatial domain. In Fig. \ref{S_SER}, we evaluate the SER of different schemes with respect to the number of satellites, given the device activity $p_{\lambda}=0.1$, pilot overhead $\rho=0.4$ and $\text{SNR}=5$ dB. As expected, the SER for cooperative schemes diminishes with an increase in satellite numbers, whereas the SER for non-cooperative schemes remains relatively unchanged. Besides, only marginal gains are observed when the number of satellites is more than four. This plateau effect likely results from the satellites' circular equidistant arrangement, which shortens inter-satellite distances and leads to highly correlated channels among adjacent satellites.
It is worth exploring the optimal satellite positioning to enhance cooperative service to ground devices in the future work. Fig. \ref{S_AER} depicts the AER achieved by different schemes as a function of the number of satellites. A similar trend to Fig. \ref{S_SER} is observed. The intuitive explanation behind this trend is that when one LEO satellite misjudges the activity of a device due to unfavorable channel conditions, observations from other satellites can greatly correct this error.

\begin{figure}
	\centering
	\captionsetup{font={small}}
	\setlength{\belowcaptionskip}{-.6cm}
	\subfigure[Performance of channel estimation.]{
		\label{Antenna_NMSE}
		\begin{minipage}{7cm}
			\includegraphics[width=\textwidth]{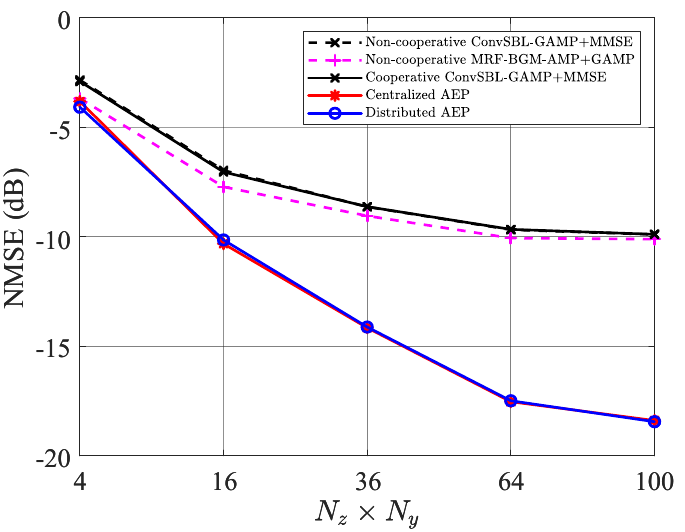} \\
			\vspace{-0.2cm}	
		\end{minipage}
	}
	\subfigure[Performance of symbol detection.]{
		\label{Antenna_SER}
		\begin{minipage}{7cm}
			\includegraphics[width=\textwidth]{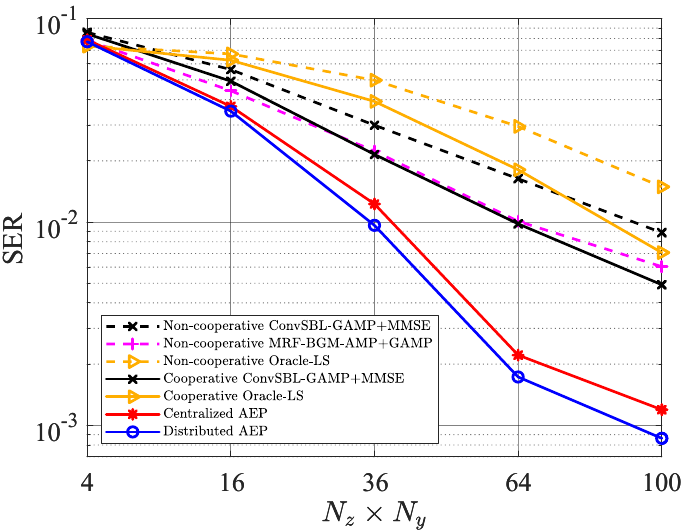} \\
			\vspace{-0.2cm}	
		\end{minipage}
	}
	\vspace{-0.2cm}
	\caption{Performance comparison among different schemes given different number of antennas, where $S_a=2$, $p_{\lambda}=0.1$, $\rho=0.4$, and $\text{SNR}=5$ dB.} 
	\vspace{0.6cm}
	\label{AntennaPer}
\end{figure}
Finally, Fig. \ref{AntennaPer} depicts the NMSE and SER as the functions of the number of antennas achieved by various algorithms, given the number of satellites $S_a=2$, device activity $p_{\lambda}=0.1$, pilot overhead $\rho=0.4$, and $\text{SNR}=5$ dB. As expected, increasing the number of receive antennas enhances the performance of all schemes. This improvement is due to higher spatial resolution at the receiver, which reduces inter-user and inter-component interference. Additionally, both the proposed centralized and distributed AEP algorithms outperform the benchmarks significantly as the number of antennas increases, highlighting their effectiveness in exploiting the sparsity of the angular domain channel.

\section{Conclusion}
\label{VI}
This paper proposed a random access scheme tailored for cooperative LEO satellite networks, leveraging OTFS to mitigate the significant Doppler effect in the TSLs. We introduced GCE-BEM to obtain a compact channel representation and then analyzed the input-output relationship of the system. Next, by exploiting the 2D block sparsity of the large antenna array, a message-passing type algorithm was designed for initial channel estimation. Then, based on the EP and CLT, we proposed both a centralized and distributed FFT-based cooperative algorithm for refining channel estimation and symbol detection of active devices, respectively. Simulation results have verified that our proposed schemes outperform existing solutions in terms of AER, NMSE and SER. Furthermore, the distributed method requires only two exchanges of soft information about data symbols within the constellation to achieve performance comparable to the centralized scheme. Future work is worth exploring the optimal satellite positioning to enhance cooperative service to ground devices.

\appendices
\section{Proof of Proposition 1}
\label{App1}
Firstly, we denote the $(l,n^{\prime})$-th element of $\mathbf{B}_{u}^{\mathrm{TD}}$ as $B_{u}^{\mathrm{TD}}[l,n^{\prime}]$, where $l=0,\dots,M-1$ and $n^{\prime}=0,\dots,N-1$. Then, substituting (\ref{r}) into (\ref{YDD}), we have
	\begin{align}
	\label{YDDscalar}
	&Y^{\text{DD}}_{n_a,u,s_a}[l,k]=\frac{1}{\sqrt{N}} \sum_{n^{\prime}=0}^{N-1}\sum_{l^{\prime} \leqslant l} h_{u,s_a}^{n_a} \left[M_{\text{cp}}+n^{\prime}M+l, l^{\prime}\right] \nonumber \\
	&\qquad \qquad \qquad \qquad \qquad \qquad \times B_{u}^{\mathrm{TD}}\left[l-l^{\prime}, n^{\prime}\right] 
	e^{-\bar{\jmath} \frac{2 \pi}{N} k n^{\prime}} \nonumber \\
	&\qquad+\frac{1}{\sqrt{N}} \sum_{n^{\prime}=0}^{N-1}\sum_{l^{\prime}>l} h_{u,s_a}^{n_a}\left[M_{\text{cp}}+n^{\prime}M+l, l^{\prime}\right]
	e^{-\bar{\jmath} \frac{2 \pi}{N} k n^{\prime}}\nonumber \\
	&\qquad \qquad \qquad \qquad \quad\times B_{u}^{\mathrm{TD}}\left[l-l^{\prime}+M, (n^{\prime}-1)_N\right]
	. 
	\end{align}
	It is observed that (\ref{YDDscalar}) is the sum of two parts, where the former one is the sum for $l^{\prime} \leqslant l$ after demodulation and the other represents the sum for $l^{\prime} > l$ after demodulation. We denote the two parts as $\tilde S_{u,s_a}^{{n_a}}[l,k]$ and $\bar S_{u,s_a}^{n_a}[l,k]$, respectively. Then, $Y^{\text{DD}}_{n_a,u,s_a}[l,k]$ can be represented as $Y^{\text{DD}}_{n_a,u,s_a}[l,k]=\tilde S_{u,s_a}^{n_a}[l,k]+\bar S_{u,s_a}^{n_a}[l,k]$. Substituting (\ref{has}) and (\ref{B}) into (\ref{YDDscalar}) and given $\omega_q = \frac{2 \pi q^{\prime}}{M N R}$ with $q^{\prime} = \left(q-\left\lceil\frac{Q}{2}\right\rceil\right)$, we can get
	\begin{align}
	&\tilde S_{u,s_a}^{n_a}[l,k] = \frac{1}{N} \sum_{q=0}^Q \sum_{l^{\prime} \leqslant l} \sum_{n^{\prime}=0}^{N-1} \sum_{k^{\prime}=0}^{N-1} 	
	c_{q, l^{\prime}, u, s_a}^{n_a}
	e^{\bar{\jmath} \frac{2 \pi q^{\prime}}{M N R} \left(M_{\text{cp}}+n^{\prime} M+l\right)}
	\nonumber \\
	&\qquad \qquad \qquad \qquad \quad \times X^{\text{DD}}_u\left[l-l^{\prime}, k^{\prime}\right]
	e^{\bar{\jmath} \frac{2 \pi}{N} (k^{\prime}-k) n^{\prime}}\nonumber \\
	&=\frac{1}{N}\sum_{q=0}^Q \sum_{l^{\prime} \leqslant l} \sum_{k^{\prime}=0}^{N-1} 
	c_{q, l^{\prime}, u, s_a}^{n_a}
	e^{\bar{\jmath} \frac{2 \pi q^{\prime}}{MNR}  (M_{\text{cp}}+l)}	
	X^{\text{DD}}_u\left[l-l^{\prime}, k^{\prime}\right] 
	\nonumber \\
	&\qquad \qquad \qquad \qquad\times \frac{1-e^{\bar{\jmath} 2 \pi\left(k^{\prime}-\left(k-\frac{q^{\prime}}{R}\right)\right)}}{1-e^{\bar{\jmath} \frac{2 \pi}{N}\left(k^{\prime}-\left(k-\frac{q^{\prime}}{R}\right)\right)}},
	\end{align}
	and 
	\begin{align}
	&\bar S_{u,s_a}^{n_a}[l,k] \nonumber \\
	&=\frac{1}{N} \sum_{q=0}^Q \sum_{l^{\prime} > l}  \sum_{k^{\prime}=0}^{N-1} 	
	c_{q, l^{\prime}, u, s_a}^{n_a}
	e^{\bar{\jmath} \frac{2 \pi q^{\prime}}{M N R} \left(M_{\text{cp}}+l\right)}
	X^{\text{DD}}_u\left[l-l^{\prime}+M, k^{\prime}\right] 
	\nonumber \\
	&\qquad \qquad \qquad \qquad \times \sum_{n^{\prime}=0}^{N-1}
	e^{\bar{\jmath} \frac{2 \pi q^{\prime}}{RN}n^{\prime}}
	e^{\bar{\jmath} \frac{2 \pi}{N} k^{\prime} (n^{\prime}-1)_N}
	e^{-\bar{\jmath} \frac{2 \pi}{N} kn^{\prime}} \nonumber \\
	&=\frac{1}{N} \sum_{q=0}^Q \sum_{l^{\prime} > l}  \sum_{k^{\prime}=0}^{N-1} 	
	c_{q, l^{\prime}, u, s_a}^{n_a}
	e^{\bar{\jmath} \frac{2 \pi q^{\prime}}{M N R} \left(M_{\text{cp}}+l\right)}
	X^{\text{DD}}_u\left[l-l^{\prime}+M, k^{\prime}\right]
	\nonumber \\
	&\qquad \qquad \qquad \qquad \times e^{-\bar{\jmath} \frac{2 \pi}{N} k^{\prime}} \frac{1-e^{\bar{\jmath} 2 \pi\left(k^{\prime}-\left(k-\frac{q^{\prime}}{R}\right)\right)}}{1-e^{\bar{\jmath} \frac{2 \pi}{N}\left(k^{\prime}-\left(k-\frac{q^{\prime}}{R}\right)\right)}}.
	\end{align} 
	Finally, by absorbing the constant $e^{\bar{\jmath} \frac{2 \pi q^{\prime}}{M N R} M_{\text{cp}}}$ into $c_{q, l^{\prime}, u, s_a}^{n_a}$ and substituting $\frac{2 \pi q^{\prime}}{M N R}$ as $\omega_q$, we get the result in Proposition 1.

\section{Message Scheduling in the MRF}
\label{App2}
With the inputs $\Delta^{f^{c^{u,s_a}}_{i,j}}_{s^{u,s_a}_{i,j}}$, we are now ready to describe the messages involved in the MRF. Recalling that the satellite adopts the 2D UPA, $\mathcal{D}_j$ can be given as $\mathcal{D}_j=\{j+N_z,j-N_z,j+1,j-1\}$. To clearly characterize
the relative position, the left, right, top, and bottom neighbors of $s_{i,j}^{u,s_a}$ are reindexed by $\{s_{i,j_{\text L}}^{u,s_a}, s_{i,j_{\text R}}^{u,s_a}, s_{i,j_{\text T}}^{u,s_a}, s_{i,j_{\text B}}^{u,s_a} \}$ corresponding to $\mathcal{D}_{j}$. 
The left, right, top, and bottom input messages of $s_{i,j}^{u,s_a}$ are denoted as $\Omega^{\text{L}^{u,s_a}}_{i,j}$, $\Omega^{\text{R}^{u,s_a}}_{i,j}$, $\Omega^{\text{T}^{u,s_a}}_{i,j}$, and $\Omega^{\text{B}^{u,s_a}}_{i,j}$. Then, the input message of $s_{i,j}^{u,s_a}$ from the left is given by
\begin{align}
\label{TSEfirst}
&\Omega^{\text{L}^{u,s_a}}_{i,j}\nonumber \\
&\propto \int_{\sim s_{i,j}^{u,s_a}}
\Delta^{f^{c^{u,s_a}}_{i,j}}_{s^{u,s_a}_{i,j}} 
\Gamma(s_{i,j}^{u,s_a}, s_{i,j_{\text L}}^{u,s_a}) 
\Psi(s_{i,j_{\text L}}^{u,s_a}) 
\prod_{\text p \in \{\text L,\text T,\text B\}} 
\Omega^{\text p^{u,s_a}}_{i,j_{\text{L}}} 
\nonumber \\
&=\xi^{\text{L}^{u,s_a}}_{i,j} \delta\left(s_{i, j}^{u,s_a}-1\right) + (1-\xi^{\text{L}^{u,s_a}}_{i,j}) \delta\left(s_{i, j}^{u,s_a}+1\right),
\end{align}
where $\xi^{\text{L}^{u,s_a}}_{i,j}$ is given by (\ref{xiupdate}).
\begin{figure*}[!t]
	\normalsize
	\setcounter{equation}{68}
	\begin{equation}
	\label{xiupdate}
	\xi^{\text{L}^{u,s_a}}_{i,j} = \frac{e^{-\gamma+\beta}
		\eta_{i,j_{\text L}}^{u,s_a}
		\prod_{\text p \in \{\text L,\text T,\text B\}} \xi^{\text p^{u,s_a}}_{i,j_{\text L}}	
		+ 
		e^{\gamma-\beta}
		(1-\eta_{i,j_{\text L}}^{u,s_a})
		\prod_{\text p \in \{\text L,\text T,\text B\}} (1-\xi^{\text p^{u,s_a}}_{i,j_{\text L}})}{(e^{\beta}+e^{-\beta})
		\left(e^{-\gamma}
		\eta_{i,j_{\text L}}^{u,s_a}
		\prod_{\text p \in \{\text L,\text T,\text B\}} \xi^{\text p^{u,s_a}}_{i,j_{\text L}}+
		e^{\gamma}
		(1-\eta_{i,j_{\text L}}^{u,s_a})
		\prod_{\text p \in \{\text L,\text T,\text B\}} (1-\xi^{\text p^{u,s_a}}_{i,j_{\text L}})\right)}.
	\end{equation}
\end{figure*}
The input messages of $s_{i,j}^{u,s_a}$ from right, top, and bottom have a similar form to $\Omega^{\text{L}^{u,s_a}}_{i,j}$. Then, the output message for $s_{i,j}^{u,s_a}$ is given by
\begin{align}
\label{zitaupdate}
&\Delta_{f^{c^{u,s_a}}_{i,j}}^{s^{u,s_a}_{i,j}} 
\propto \Psi(s_{i,j_{\text L}}^{u,s_a}) 
\prod_{\text p \in \{\text L,\text R,\text T,\text B\}}
\Omega^{\text p^{u,s_a}}_{i,j} \nonumber \\
&= \zeta_{i,j}^{u,s_a} \delta\left(s_{i, j}^{u, l^{\prime}}-1\right) +
(1 - \zeta_{i,j}^{u,s_a}) \delta\left(s_{i, j}^{u, l^{\prime}}+1\right),
\end{align}
where 
\begin{align}
\label{zeta}
	\zeta_{i,j}^{u,s_a}\! = \!\frac{e^{-\gamma} \prod_{\text p\in\{\text L, \text R, \text T, \text B\}} \xi^{\text p^{u,s_a}}_{i,j}}{e^{-\gamma} \prod_{\text p\in\{\text L, \text R, \text T, \text B\}} \xi^{p^{u,s_a}}_{i,j} + e^{\gamma} \prod_{\text p\in\{\text L, \text R, \text T, \text B\}} (1-\xi^{\text p^{u,s_a}}_{i,j})}.
\end{align}

\section{Message Scheduling in the IC Module}
\label{App3}
In this appendix, we investigate the message passing in the IC module. With the input message $\Delta_{f^{\text{P}}_{q, u, n_a, s_a}}^{\mathbf x^{\text F}_{u}} = \mathcal{CN}(\mathbf x^{\text F}_{u}\!\!\mid \overrightarrow{\mathbf{x}}^{\text F}_{q, u, n_a, s_a}, \operatorname{diag}(\overrightarrow{\bm{\tau}}^{x_{\text F}}_{q, u, n_a, s_a}))$ and $\Delta_{f^{\text{P}}_{q, u, n_a, s_a}}^{\mathbf c_{q,u,n_a, s_a}^{\text F}}=\mathcal{CN}(\mathbf c^{\text F}_{q,u,n_a, s_a}\mid \overleftarrow{\mathbf{c}}^{\text F}_{q,u,n_a, s_a}, \operatorname{diag}(\overleftarrow{\bm{\tau}}^{c_{\text F}}_{q,u,n_a, s_a}))$, the message $\Delta^{f^{\text{P}}_{q, u, n_a, s_a}}_{\mathbf d^{\text{P}}_{q, u, n_a, s_a}}$ passed into the IC module can be approximated as a Gaussian distribution, i.e., $\Delta^{f^{\text{P}}_{q, u, n_a, s_a}}_{\mathbf d^{\text{P}}_{q, u, n_a, s_a}} = \mathcal{CN}(\mathbf d^{\text{P}}_{q, u, n_a, s_a} \mid \overleftarrow{\mathbf d}^{\text{P}}_{q, u, n_a, s_a}, \operatorname{diag}(\overleftarrow{{\bm{\tau}}}_{q, u, n_a, s_a}^{d_{\text{P}}}))$, where $\overleftarrow{\mathbf d}^{\text{P}}_{q, u, n_a, s_a}=\overrightarrow{\mathbf{x}}^{\text F}_{u} \odot \overleftarrow{\mathbf{c}}^{\text F}_{q,u,n_a, s_a}$ and $\overleftarrow{{\bm{\tau}}}_{q, u, n_a, s_a}^{d_{\text{P}}}=\overrightarrow{\bm{\tau}}^{x_{\text F}}_{u} \odot \overleftarrow{\bm{\tau}}^{c_{\text F}}_{q,u,n_a, s_a} + \overrightarrow{\mathbf{x}}^{\text F}_{u} \odot \overrightarrow{\mathbf{x}}^{\text F^*}_{u} \odot \overleftarrow{\bm{\tau}}^{c_{\text F}}_{q,u,n_a, s_a} + \overleftarrow{\mathbf{c}}^{\text F}_{q,u,n_a, s_a} \odot \overleftarrow{\mathbf{c}}^{\text F^*}_{q,u,n_a, s_a} \odot \overrightarrow{\bm{\tau}}^{x_{\text F}}_{u}$.
Since $\Delta^{\mathbf d^{\text{P}}_{q, u, n_a, s_a}}_{f^{\text W}_{q,n_a,s_a}}=\Delta^{f^{\text{P}}_{q, u, n_a, s_a}}_{\mathbf d^{\text{P}}_{q, u, n_a, s_a}}$, the message from factor node $f^{\text W}_{q,n_a,s_a}$ to the variable $\mathbf d^{\text{W}}_{q, n_a, s_a}$ is given as 
\begin{align}
\label{ICstart}
&\Delta^{f^{\text W}_{q,n_a,s_a}}_{\mathbf d^{\text{W}}_{q, n_a, s_a}} \nonumber \\ 
&\propto
\frac{\operatorname{Proj}\left[\Delta_{f^{\text W}_{q,n_a,s_a}}^{\mathbf d^{\text{W}}_{q, n_a, s_a}}
	\int_{\sim \mathbf d^{\text{W}}_{q, n_a, s_a}}
	f^{\text W}_{q,n_a,s_a}\prod_u \Delta^{\mathbf d^{\text{P}}_{q, u, n_a, s_a}}_{f^{\text W}_{q,n_a,s_a}}\right]}{\Delta_{f^{\text W}_{q,n_a,s_a}}^{\mathbf d^{\text{W}}_{q, n_a, s_a}}} \nonumber \\
&\overset{(a)}{\propto}
\mathcal{CN}(\mathbf d^{\text{W}}_{q, u, n_a, s_a} \mid \overleftarrow{\mathbf d}^{\text{W}}_{q, u, n_a, s_a}, \operatorname{diag}(\overleftarrow{{\bm{\tau}}}_{q, u, n_a, s_a}^{d_{\text{W}}})),
\end{align}
where $(a)$ is due to CLT, $\overleftarrow{\mathbf d}^{\text{W}}_{q, u, n_a, s_a}=\sum_u \overleftarrow{\mathbf d}^{\text{P}}_{q, u, n_a, s_a}$, and $\overleftarrow{{\bm{\tau}}}_{q, u, n_a, s_a}^{d_{\text{W}}}=\sum_u \overleftarrow{{\bm{\tau}}}_{q, u, n_a, s_a}^{d_{\text{P}}}$. Since $\mathbf d^{\text{W}}_{q, n_a, s_a}$ is transformed to $\mathbf d^{\text F}_{q, n_a, s_a}$ by the DFT matrix and we have $\Delta^{\mathbf d^{\text{W}}_{q, n_a, s_a}}_{f^{\mathbf d_{\text F}}_{q,n_a,s_a}}=\Delta^{f^{\text W}_{q,n_a,s_a}}_{\mathbf d^{\text{W}}_{q, n_a, s_a}}$, the message $\Delta_{\mathbf d^{\text F}_{q, n_a, s_a}}^{f^{\mathbf d_{\text F}}_{q,n_a,s_a}}$ can be obtained as 
$\Delta_{\mathbf d^{\text F}_{q, n_a, s_a}}^{f^{\mathbf d_{\text F}}_{q,n_a,s_a}}=
\mathcal{CN}(\mathbf d^{\text F}_{q, n_a, s_a} \mid \overleftarrow{\mathbf d}^{\text{F}}_{q, n_a, s_a}, \overleftarrow{{\tau}}_{q, n_a, s_a}^{d_{\text{F}}}\mathbf I_{MN})$, where $\overleftarrow{\mathbf d}^{\text{F}}_{q, n_a, s_a}=\mathbf{F}_{MN}^{\mathrm{H}} \overleftarrow{\mathbf d}^{\text{W}}_{q, u, n_a, s_a}$, $\overleftarrow{{\tau}}_{q, n_a, s_a}^{d_{\text{F}}}=<\overleftarrow{{\bm{\tau}}}_{q, u, n_a, s_a}^{d_{\text{W}}}>$, and $<\cdot>$ is the empirical averaging operation. Recalling that $\mathbf{b}_q$ represents the $q$-th basis function of BEM, and hence $\operatorname{diag}(\mathbf{b}_q)$ is a unitary matrix. The message $\Delta^{f^{\text B}_{q,n_a,s_a}}_{\mathbf d^{\text{B}}_{q, n_a, s_a}}$ can be computed as $\Delta^{f^{\text B}_{q,n_a,s_a}}_{\mathbf d^{\text{B}}_{q, n_a, s_a}}=\mathcal{CN}(\mathbf d^{\text{B}}_{q, n_a, s_a} \mid \overleftarrow{\mathbf d}^{\text{B}}_{q, n_a, s_a}, \overleftarrow{{\tau}}_{q, n_a, s_a}^{d_{\text{B}}}\mathbf I_{MN})$, where $\overleftarrow{\mathbf d}^{\text{B}}_{q, n_a, s_a}=\mathbf{b}_q \odot \overleftarrow{\mathbf d}^{\text{F}}_{q, n_a, s_a}$ and $\overleftarrow{{\tau}}_{q, n_a, s_a}^{d_{\text{B}}}=\overleftarrow{{\tau}}_{q, n_a, s_a}^{d_{\text{F}}}$. Then, the message keep passing from right to left in the factor graph and we have $\Delta_{f^{\mathbf g}_{n_a,s_a}}^{\mathbf d^{\text{B}}_{q, n_a, s_a}} = \Delta^{f^{\text B}_{q,n_a,s_a}}_{\mathbf d^{\text{B}}_{q, n_a, s_a}}$. At this point, the message start to pass in the opposite direction. According to the CLT, the output message of $f^{\mathbf g}_{n_a,s_a}$ is derived as
\begin{align}
\label{ICC1}
&\Delta^{f^{\mathbf g}_{n_a,s_a}}_{\mathbf d^{\text{B}}_{q, n_a, s_a}} \nonumber \\
&\propto
\frac{\operatorname{Proj}\left[\Delta_{f^{\mathbf g}_{n_a,s_a}}^{\mathbf d^{\text{B}}_{q, n_a, s_a}}
	\int_{\sim \mathbf d^{\text{B}}_{q, n_a, s_a}} f^{\mathbf g}_{n_a,s_a} \Delta_{f^{\mathbf g}_{n_a,s_a}}^{\mathbf g_{n_a, s_a}} \prod_{\bar q \neq q}  \Delta^{f^{\text B}_{\bar q,n_a,s_a}}_{\mathbf d^{\text{B}}_{\bar q, n_a, s_a}}\right]}{\Delta_{f^{\mathbf g}_{n_a,s_a}}^{\mathbf d^{\text{B}}_{q, n_a, s_a}}}
\nonumber \\
&\propto \mathcal{CN}(\mathbf d^{\text{B}}_{q, n_a, s_a} \mid \overrightarrow{\mathbf d}^{\text{B}}_{q, n_a, s_a}, \overrightarrow{{\tau}}_{q, n_a, s_a}^{d_{\text{B}}}\mathbf I_{MN}),
\end{align}
where $\overrightarrow{\mathbf d}^{\text{B}}_{q, n_a, s_a}=\mathbf{y}_{n_a,s_a}^{\prime} -\sum_{\bar q \neq q} \overleftarrow{\mathbf d}^{\text{B}}_{\bar q, n_a, s_a}$ and $\overrightarrow{{\tau}}_{q, n_a, s_a}^{d_{\text{B}}}=\sigma^2_{s_a} + \sum_{\bar q \neq q} \overleftarrow{{\tau}}_{\bar q, n_a, s_a}^{d_{\text{B}}}$. It can be observed that this step is similar with the SIC and eliminates the inter-component interference in BEM using the soft estimates. 
By the symmetry, we can follow the similar steps to derive the message in the opposite direction, and the message $\Delta^{f^{\text B}_{q,n_a,s_a}}_{\mathbf d^{\text F}_{q, n_a, s_a}}=\mathcal{CN}(\mathbf d^{\text F}_{q, n_a, s_a} \mid \overrightarrow{\mathbf d}^{\text{F}}_{q, n_a, s_a}, \overrightarrow{{\tau}}_{q, n_a, s_a}^{d_{\text{F}}}\mathbf I_{MN})$ with $\overrightarrow{\mathbf d}^{\text{F}}_{q, n_a, s_a}=\mathbf{b}_q^{*} \odot \overrightarrow{\mathbf d}^{\text{B}}_{q, n_a, s_a}$ and $\overrightarrow{{\tau}}_{q, n_a, s_a}^{d_{\text{F}}}=\overrightarrow{{\tau}}_{q, n_a, s_a}^{d_{\text{B}}}$; the message $\Delta_{\mathbf d^{\text{W}}_{q, n_a, s_a}}^{f^{\mathbf d_{\text F}}_{q,n_a,s_a}}=
\mathcal{CN}(\mathbf d^{\text{W}}_{q, n_a, s_a} \mid \overrightarrow{\mathbf d}^{\text{W}}_{q, n_a, s_a}, \overrightarrow{{\tau}}_{q, n_a, s_a}^{d_{\text{W}}}\mathbf I_{MN})$, where $\overrightarrow{\mathbf d}^{\text{W}}_{q, n_a, s_a}=\mathbf{F}_{MN} \overrightarrow{\mathbf d}^{\text{F}}_{q, n_a, s_a}$ and $\overleftarrow{{\tau}}_{q, n_a, s_a}^{d_{\text{W}}}=\overrightarrow{{\tau}}_{q, n_a, s_a}^{d_{\text{F}}}$. Similar to (\ref{ICC1}), the output message of $f^{\text W}_{q,n_a,s_a}$ is obtained by combining relevant messages, i.e.,
\begin{align}
\label{ICend}
&\Delta_{\mathbf d^{\text{P}}_{q, u, n_a, s_a}}^{f^{\text W}_{q,n_a,s_a}} \propto \nonumber \\
&
\scriptstyle \frac{\operatorname{Proj}\left[\Delta^{\mathbf d^{\text{P}}_{q, u, n_a, s_a}}_{f^{\text W}_{q,n_a,s_a}}
	\int_{\sim \mathbf d^{\text{P}}_{q, u, n_a, s_a}} f^{\text W}_{q,n_a,s_a} \Delta_{f^{\text W}_{q,n_a,s_a}}^{\mathbf d^{\text{W}}_{q, n_a, s_a}} \prod_{\bar u \neq u}  \Delta^{\mathbf d^{\text{P}}_{q, \bar u, n_a, s_a}}_{f^{\text W}_{q,n_a,s_a}}\right]}{\Delta^{\mathbf d^{\text{P}}_{q, u, n_a, s_a}}_{f^{\text W}_{q,n_a,s_a}}}
\nonumber \\
&\propto \mathcal{CN}(\mathbf d^{\text{P}}_{q, u, n_a, s_a} \mid \overrightarrow{\mathbf d}^{\text{P}}_{q, u, n_a, s_a}, \operatorname{diag}(\overrightarrow{\bm{\tau}}_{q, u, n_a, s_a}^{d_{\text{P}}})),
\end{align}
where $\overrightarrow{\mathbf d}^{\text{P}}_{q, u, n_a, s_a} = \overrightarrow{\mathbf d}^{\text{W}}_{q, n_a, s_a}-\sum_{\bar u \neq u} \overleftarrow{\mathbf d}^{\text{P}}_{q, \bar u, n_a, s_a}$ and $\overrightarrow{\bm{\tau}}_{q, u, n_a, s_a}^{d_{\text{P}}}=\overleftarrow{{\tau}}_{q, n_a, s_a}^{d_{\text{W}}} \mathbf 1_{MN} + \sum_{\bar u \neq u} \overleftarrow{{\bm{\tau}}}_{q, \bar u, n_a, s_a}^{d_{\text{P}}}$. Note that the inter-user interference is canceled in this step. 
Similar to \cite{c21}, we next resort to the mean-filed approximation to provide the likelihoods for $\mathbf c^{\text F}_{q,u,n_a, s_a}$ and $\mathbf x^{\text F}_{u}$. Given the posterior distribution $\Delta_{\mathbf x_u}^{\text F}=\mathcal{CN}(\mathbf x^{\text F}_{u}\mid \hat{\mathbf{x}}^{\text F}_{u}, \operatorname{diag}(\bm{\tau}^{x_{\text F}}_{u}))$ and $\Delta_{\mathbf c_{q,u,n_a, s_a}}^{\text F}=\mathcal{CN}(\mathbf c^{\text F}_{q,u,n_a, s_a}\mid \hat{\mathbf{c}}^{\text F}_{q,u,n_a, s_a}, \operatorname{diag}(\bm{\tau}^{c_{\text F}}_{q,u,n_a, s_a}))$, respectively, we can get
\begin{align}
\label{ICx}
&\Delta^{f^{\text{P}}_{q, u, n_a, s_a}}_{\mathbf x^{\text F}_{u}} 
\propto
\exp\int_{\mathbf c_{q,u,n_a, s_a}^{\text F}} 
\Delta_{\mathbf c_{q,u,n_a, s_a}}^{\text F} \times \nonumber \\
&\log \mathcal{CN}(\mathbf c^{\text F}_{q,u,n_a, s_a} \odot \mathbf x_u^{\text F} \mid \overrightarrow{\mathbf d}^{\text{P}}_{q, u, n_a, s_a}, \operatorname{diag}(\overrightarrow{\bm{\tau}}_{q, u, n_a, s_a}^{d_{\text{P}}})) \nonumber \\
&\propto
\mathcal{CN}(\mathbf x^{\text F}_{u}\mid \overleftarrow{\mathbf{x}}^{\text F}_{q, u, n_a, s_a}, \operatorname{diag}(\overleftarrow{\bm{\tau}}^{x_{\text F}}_{q, u, n_a, s_a})),
\end{align}
where $\overleftarrow{\mathbf{x}}^{\text F}_{q, u, n_a, s_a}=\frac{\overrightarrow{\mathbf d}^{\text{P}}_{q, u, n_a, s_a} \odot \hat{\mathbf{c}}^{{\text F}^*}_{q,u,n_a, s_a}}{\hat{\mathbf{c}}^{\text F}_{q,u,n_a, s_a}\odot \hat{\mathbf{c}}^{{\text F}^*}_{q,u,n_a, s_a}+\bm{\tau}^{c_{\text F}}_{q,u,n_a, s_a}}$ and $\overleftarrow{\bm{\tau}}^{x_{\text F}}_{q, u, n_a, s_a}=\frac{\overrightarrow{\bm{\tau}}_{q, u, n_a, s_a}^{d_{\text{P}}}}{\hat{\mathbf{c}}^{\text F}_{q,u,n_a, s_a}\odot \hat{\mathbf{c}}^{{\text F}^*}_{q,u,n_a, s_a}+\bm{\tau}^{c_{\text F}}_{q,u,n_a, s_a}}$. By the symmetry, we also have 
\begin{align}
\label{ICc}
\Delta^{f^{\text{P}}_{q, u, n_a, s_a}}_{\mathbf c_{q,u,n_a, s_a}^{\text F}}=\mathcal{CN}(\mathbf c^{\text F}_{q,u,n_a, s_a}\mid \overrightarrow{\mathbf{c}}^{\text F}_{q,u,n_a, s_a}, \operatorname{diag}(\overrightarrow{\bm{\tau}}^{c_{\text F}}_{q,u,n_a, s_a})),
\end{align}
where $\overrightarrow{\mathbf{c}}^{\text F}_{q,u,n_a, s_a}=
\frac{\overrightarrow{\mathbf d}^{\text{P}}_{q, u, n_a, s_a} \odot \hat{\mathbf{x}}^{{\text F}^*}_{u}}{\hat{\mathbf{x}}^{\text F}_{u} \odot \hat{\mathbf{x}}^{{\text F}^*}_{u}+\bm{\tau}^{x_{\text F}}_{u}}$ 
and $\overrightarrow{\bm{\tau}}^{c_{\text F}}_{q,u,n_a, s_a}=\frac{\overrightarrow{\bm{\tau}}_{q, u, n_a, s_a}^{d_{\text{P}}}}{\hat{\mathbf{x}}^{\text F}_{u} \odot \hat{\mathbf{x}}^{{\text F}^*}_{u} +\bm{\tau}^{x_{\text F}}_{u}}$.

\bibliographystyle{IEEEtran}
\bibliography{bibfile}
\end{document}